\documentclass[prl,reprint,amssymb,superscriptaddress]{revtex4-2}

\usepackage{xcolor}

\usepackage{bm}

\usepackage{amsmath}  
\usepackage{amsfonts}
\usepackage{graphicx}
\usepackage{hyperref}
\usepackage{comment}
\usepackage[normalem]{ulem}
\hypersetup{
	colorlinks=true,
	citecolor=blue,
	linkcolor=red,
	urlcolor = blue
}

\begin{document}
	
	\title{Deciphering the origin of spin current in spintronic terahertz emitters and its imprint on their electromagnetic radiation via time-dependent density functional theory}

	\author{Ali Kefayati}
	\affiliation{Department of Physics and Astronomy, University of Delaware, Newark, DE 19716, USA}
	\author{Yafei Ren}
	\affiliation{Department of Physics and Astronomy, University of Delaware, Newark, DE 19716, USA}
	\author{M. Benjamin Jungfleisch}
	\affiliation{Department of Physics and Astronomy, University of Delaware, Newark, DE 19716, USA}
	\author{Lars Gundlach}
	\affiliation{Department of Physics and Astronomy, University of Delaware, Newark, DE 19716, USA}
	\affiliation{Department of Chemistry and Biochemistry, University of Delaware, Newark, DE 19716, USA}
	\author{John Q. Xiao}
	\affiliation{Department of Physics and Astronomy, University of Delaware, Newark, DE 19716, USA}
	\author{Branislav K. Nikoli\'c}
	\email{bnikolic@udel.edu}
	\affiliation{Department of Physics and Astronomy, University of Delaware, Newark, DE 19716, USA}

	\begin{abstract}
		Spin current flowing between femtosecond laser pulse (fsLP)-driven  ferromagnetic metal and adjacent normal metal (NM) hosting strong spin-orbit coupling is invariably invoked to explain terahertz (THz) radiation believed to be emitted solely by NM layer. Despite being such a central concept, the microscopic origin of interlayer spin current remains vague. It is also unclear if it directly imprints its time-dependence onto emitted THz  radiation because far-field radiation can only be emitted by time-dependent charge current, while a potentially large number of complex spin-to-charge conversion mechanisms can obscure direct relation between interlayer spin current and converted charge current within NM layer. Here, we employ recently developed [A. Kefayati {\em et al.}, Phys. Rev. Lett. {\bf 133}, 136704 (2024)]  time-dependent density functional theory plus Jefimenko equations approach to  extract spin current between Co and NM=Pt or NM=W layer where Co is driven by fsLP responsible for its demagnetization, i.e., shrinking of its  magnetization vector, $M^y(t)/M^y(t=0)<1$. By comparing time dependence of  spin current with those of other relevant quantities, we find that: ({\em i}) spin current is generated by demagnetization dynamics because it {\em follows} closely $dM^y/dt$, thus it is an example of quantum pumping phenomenon that cannot be captured by phenomenological notions (such as ``spin voltage'') and related semiclassical transport theories;  ({\em ii}) time dependence of pumped spin current {\em does not follow} closely  that of charge current emerging within NM layer via spin-to-charge conversion mechanisms; ({\em iii}) THz emission can be governed by {\em both} charge current (i.e., its time derivative entering the Jefimenko equations) within Co layer or NM layer, 
		but in different times frames. We also unravel a special case of NM=W where spin-to-charge conversion by the inverse spin Hall effect and its contribution to THz emission is 
		suppressed, despite large spin Hall angle of W, because of localization of excited electrons onto the outer unfilled $d$-orbitals of W. 
	\end{abstract}
	
	\maketitle

\begin{figure}[h!]
		\centering
		\includegraphics[width = \linewidth]{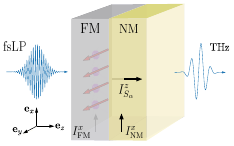}
		\caption{Schematic view of FM/NM bilayer employed~\cite{Seifert2016,Seifert2023,Wu2017,Zhang2017b,Rouzegar2022,Liu2021} for spintronic THz emitters  where fsLP irradiates FM layer, such as Co in our study, and NM layer hosts strong SOC, such as Pt or W we select in this study for comparison with each other as their spin Hall angle has opposite sign~\cite{Seifert2023}. The thickness of Co layer is three monolayers, while the thickness of NM layers is  four monolayers. The local magnetization (red arrows) of Co  layer is along the $y$-axis in equilibrium, and it remains so (i.e., $M^x \approx M^z \approx 0$) during demagnetization in nonequilibrium while only  {\em shrinking} its length. Total spin current from FM to NM layer is denoted by $I_{S_\alpha}^z$ and total charge currents within FM or NM layers are denoted by $I^x_\mathrm{FM}$ and  $I^x_\mathrm{NM}$, respectively.}
		\label{fig:fig1}
	\end{figure}

	The spin current (denoted as $I^z_{S_\alpha}$ in Fig.~\ref{fig:fig1}) flowing from femtosecond laser pulse (fsLP)-driven ferromagnetic metal (FM) layer toward adjacent  normal metal (NM) within FM/NM bilayers is one of the {\em central} concept~\cite{Seifert2016,Seifert2023,Wu2017,Zhang2017b,Zhang2017b,Rouzegar2022,Jechumtal2024,Liu2021} in the field of ultrafast magnetism~\cite{Beaurepaire1996,Kirilyuk2010,Scheid2022,Chen2025}. It is always assumed that such current is efficiently converted into charge current (denoted as $I^x_\mathrm{NM}$ in Fig.~\ref{fig:fig1}) via either the inverse spin Hall effect (ISHE)~\cite{Saitoh2006} in the bulk of NM layer or other spin-orbit coupling (SOC)-driven mechanisms at FM/NM interfaces~\cite{Jungfleisch2018a, Gueckstock2021, Wang2023} due to the strong SOC in NM layer that is usually heavy transition metals (such as Pt, W, and Ta). It is, therefore, believed~\cite{Lu2020} that optimizing interlayer spin current and spin-to-charge conversion mechanisms~\cite{Mahfouzi2014a} can enhance terahertz (THz) emission by charge current within the NM layer arising from spin-to-charge conversion.

	However, the microscopic origin of interlayer spin current vector $\big(I_{S_x}^z(t),I_{S_y}^z(t),I_{S_z}^z(t)\big)$  remains unknown. Instead, interpretation of experiments is phenomenological and relies on intuitive notions~\footnote[1]{Assuming that spin accumulation alone allows one to describe demagnetization and concomitant processes like spin current generation is equivalent to using only diagonal elements of time-dependent nonequilibrium density matrix. In contrast, recent construction~\cite{Mrudul2024} of such density matrix from TDDFT calculations shows significance of its off-diagonal elements (when represented in the basis of  KS orbitals in the GS).} like ``spin voltage''~\cite{Buehlmann2020,Rouzegar2022,Seifert2023}, as a difference between nonequilibrium spin-dependent chemical potentials~\footnote[2]{Note that  defining the distribution function of electrons in far-from-equilibrium quantum systems is required to extract meaningful effective spin-dependent chemical potentials from it and then define ``spin voltage'' as their difference. However, this is a daunting task from a rigorous viewpoint. This is because  distribution function of equilibrium quantum statistical mechanics [where it is justified by diagonal form of density matrix in the basis of energy eigenstates, such as from grand canonical density matrix  \mbox{$\hat{\rho}_\mathrm{eq} = \sum_n f(E_n)|E_n\rangle \langle E_n|$} we read off standard  {\em thermal} Fermi-Dirac distribution $f(E_n)$] or   semiclassical nonequilibrium theories (such as the Boltzmann equation, sometimes used to model ultrafast demagnetization while requiring a host of additional assumptions~\cite{Nenno2018,Nenno2019,Hurst2018})  is replaced in nonequilibrium quantum statistical mechanics with single-time-dependent density matrix or its generalization in the form of  two-times-dependent lesser Keldysh Green's functions  (GF)~\cite{Stefanucci2025}. Even when the distribution function is extracted via some  ansatz 
		(typically by assuming that Keldysh  GFs satisfy equilibrium-like relations even in  nonequilibrium), it typically has a {\em nonthermal} form that cannot be described by a single effective chemical potential and single effective temperature~\cite{Kuenzel2024,Takei2019}.}, which then drives spin current according to semiclassical transport theories~\cite{Battiato2010,Nenno2018,Nenno2019,Hurst2018,Beens2022a}. Beside the   difficulties~\footnotemark[2] in  rigorously defining distribution function in far-from-equilibrium quantum systems and extracting meaningful effective chemical potential from it~\cite{Kuenzel2024,Takei2019}, 
	these theories also do not explain why frequency spectrum of spin-to-charge converted current~\cite{Seifert2016} contains features in the THz range or the role played by demagnetization dynamics for such features because spin voltage can be nonzero due to fsLP and magnetism within FM layer even if its magnetization is not changing at all (as is the case if its SOC is artificially turned off~\cite{Krieger2015,Zhang2000a}). 
	Furthermore, another standard notion~\cite{Seifert2016,Seifert2023,Wu2017,Zhang2017b,Rouzegar2022,Jechumtal2024,Liu2021} is  that the temporal profile of charge current $I^x_\mathrm{NM} (t)$ flowing within a NM layer is in  {\em one-to-one} correspondence with the temporal profile of $I_{S_y}^z(t)$
	\begin{equation}\label{eq:ishe}
		I^x_\mathrm{NM} (t) \equiv  \theta_\mathrm{SH} I_{S_y}^z(t),
	\end{equation}
	where $\theta_\mathrm{SH}$ is the spin Hall angle~\cite{Sinova2015}. Equation~\eqref{eq:ishe} relies on assumptions~\cite{Jechumtal2024}  
	that $I_{S_y}^z(t)$ is fully absorbed inside NM layer and converted (with efficiency specified by $\theta_\mathrm{SH}$) into $I^x_\mathrm{NM} (t)$, which is  then the {\em sole} source of outgoing THz signal observed in the far-field (FF) region. These assumptions also provide the foundation for  THz emission  spectroscopy~\cite{Jechumtal2024} which {\em indirectly} extracts  the temporal profile of ultrafast spin currents, triggered by fsLP excitation of thin FM films, from the time-dependence of {\em directly} measured THz signal (see Fig. 2 in Ref.~\cite{Jechumtal2024} for an example of this procedure). That Eq.~\eqref{eq:ishe} might not be warranted has been suggested by   recent experiments~\cite{Gorchon2022} pointing to a potentially large number of complex spin-to-charge conversion mechanisms and high sensitivity to changes in the optical properties, as well as charge density equilibration~\cite{Schmidt2023},  that  can obscure simplistic one-to-one correspondence assumed in Eq.~\eqref{eq:ishe}. 

	\begin{figure}
		\centering
		\includegraphics[scale=0.8]{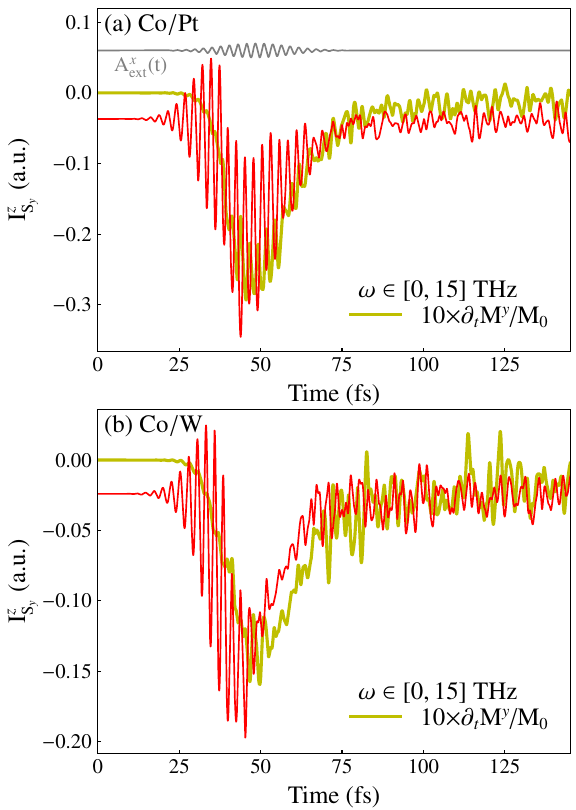}
		\caption{Time-dependence of interlayer spin current $I^z_{S_y}$ (red curves) in fsLP-driven (a) Co/Pt and (b) Co/W bilayers is compared with that of time derivative of magnetization $\partial_t M^y/M_0$ (yellow curves), demonstrating a causal connection where the latter generates (or ``pumps''~\cite{VarelaManjarres2024,Freimuth2017}) the former. Both  panels are obtained by using {\em time-domain}  filtering~\cite{Cohen2014,explaintdf}, which eventually produces a real-time signal with Fourier components within $\omega \in [0,15]$ THz range. Note that $M^y(t)/M_0$ drops (not shown explicitly) from 1 to a minimum  of $\simeq 0.5$ vs. $\simeq 0.8$ in Co/W vs. Co/Pt. Gray  curve in (a) depicts vector potential $A^x_\mathrm{ext}(t)$ of fsLP in Eq.~\eqref{eq:tddft}, and a.u. stands for atomic units.}
		\label{fig:fig2}
	\end{figure}
	\begin{figure*}
		\centering
		\includegraphics[width = \linewidth]{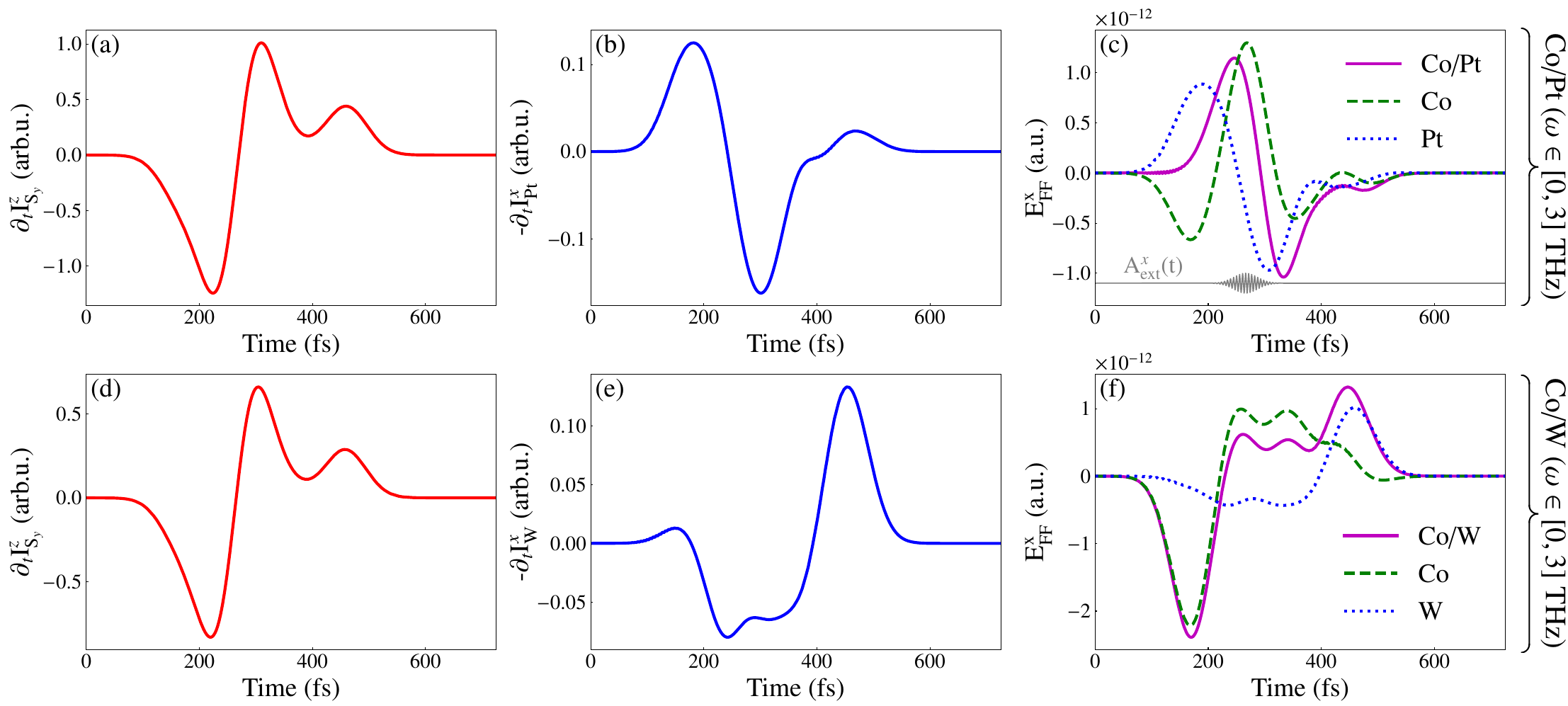}
		\caption{Time dependence of time-derivative of: (a) interlayer spin current  in Co/Pt bilayer; and (b) charge current in  Pt layer of such bilayer. Panel (c) shows the $x$-component of electric field of THz radiation  emitted by charge currents  confined within Co (green) or Pt [blue,  originating from blue curve in panel (b) via Eq.~\eqref{eq:eff}] layer, as well their sum determining the  {\em total} THz signal from Co/Pt bilayer.  Panels (d)--(f) are counterparts of panels (a)--(c) for Co/W bilayer. All six panels are obtained by using {\em time-domain}  filtering~\cite{Cohen2014,explaintdf}, which eventually produces a real-time signal with Fourier components within $\omega \in [0,3]$  THz range. Gray curve in (c) depicts vector potential $A^x_\mathrm{ext}(t)$ of fsLP in Eq.~\eqref{eq:tddft}. Note that $[0,3]$  THz range we select for filtering is often scanned  in experiments~\cite{Zhang2017b} on systems containing ultrafast demagnetizing Co layer.}
		\label{fig:fig3}
	\end{figure*}

	In this Letter, we employ time-dependent density functional theory  (TDDFT)---which has provided some of the most detailed microscopic insights~\cite{Krieger2015,Shokeen2017,Chen2019a,Pellegrini2022,Mrudul2024} into demagnetization mechanisms, and which has recently been  extended~\cite{Kefayati2024} into TDDFT+Jefimenko approach for computation of currents and the ensuing electromagnetic (EM) radiation---to analyze fsLP-driven Co/Pt and Co/W bilayers.  This approach allows us to  extract, from first principles that {\em do not make} any assumptions about the underlying physics or the system under investigation,  time-dependence of: demagnetization dynamics $M^y(t)/M_0<1$ [while $M^x(t) \approx M^z (t) \approx 0$] and its time derivative $\partial_t M^y$ [where $\partial_t \equiv \partial/\partial_t$ and $M_0 \equiv M^y(t=0)$] for magnetization in equilibrium pointing along the $y$-axis in Fig.~\ref{fig:fig1}) within Co layer [Fig.~\ref{fig:fig2}]; spin current from Co toward Pt or W layer [Fig.~\ref{fig:fig2}]; charge current flowing within Pt or W layer [Figs.~\ref{fig:fig3}(b),(e)]; and electric field $E^x_\mathrm{FF}(t)$ of THz emission in the FF region [Figs.~\ref{fig:fig3}(c),(f)]. It also makes it possible to analyze how total real-time THz signal from FM/NM bilayer emerges from contributions generated by FM (never  considered in thus far  analyzed experiments~\cite{Seifert2016,Seifert2023,Zhang2017b,Wu2017,Rouzegar2022,Jechumtal2024,Liu2021}) vs. NM layer. Note that the total signal is not necessarily the result of two contributions enhancing each other, as they can have different phases, thereby adding negative to positive lobe [as in Fig.~\ref{fig:fig3}(c)]. 
	
	We commence by examining in Fig.~\ref{fig:fig2}   the possible causal connection between $\partial_t M^y$  and interlayer spin current 
	\begin{equation}\label{eq:spincurrent}
		I_{S_\alpha}^z(t) = \int\limits_\mathrm{FM/NM} d^3\, r j_{S_\alpha}^z(\mathbf{r},t), 
	\end{equation}
	transporting spins $S_\alpha$ along the $z$-axis [see coordinate system in Fig.~\ref{fig:fig1}], which is obtained by integrating one component $j_{S_\alpha}^z(\mathbf{r},t)$ of $3 \times 3$ spin current density tensor~\cite{Dejean2022} over the whole FM/NM bilayer. Note that we use the same units for spin and charge current, where spin current $I_{S_y}^z = I_{\uparrow}^z -I_{\downarrow}^z$ can be understood as the difference of the respective spin-resolved charge currents, $I_{\uparrow}^z$ and $I_{\downarrow}^z$, with $\uparrow$, $\downarrow$ pointing along the $y$-axis. The overlapping time-dependences of $I_{S_y}^z(t)$ and $\partial_t M^y$ demonstrate that $dM^y/dt$ acts as the {\em principal mechanism} generating spin current.  This is quite analogous to spin  pumping~\cite{Tserkovnyak2005} 
	by a {\em different} type of time-dependence of magnetization, namely, its precession within FM layer of FM/NM bilayers where interlayer spin current is given by \mbox{$(I_{S_x}^z,I_{S_y}^z,I_{S_z}^z) \propto \mathbf{M} \times \partial_t \mathbf{M}$}. Here,   $\mathbf{M}(t)$ precesses as a vector of {\em fixed} length at {\em much smaller} (typically GHz) and {\em single} frequency due to microwaves being absorbed under the ferromagnetic resonance conditions~\cite{Saitoh2006, Ando2014a}. In contrast to such  precessing motion of magnetization (and absence of any demagnetization), the shrinking  magnetization vector due to high frequency fsLP   contains~\cite{Mrudul2024,GarciaGaitan2025a} a continuum of much higher frequencies in its fast-Fourier transform. Nevertheless, in both of these ``low-frequency''~\cite{Tserkovnyak2005} and ``high-frequency''~\cite{Mrudul2024,GarciaGaitan2025a} phenomena leading to $\mathbf{M}(t)$ as nonequilibrium drive for the quantum subsystem of electrons, it is time-dependence of quantum system~\cite{Citro2023, Brouwer1998}---periodic~\cite{Citro2023,Brouwer1998,Tserkovnyak2005,VarelaManjarres2023} in the case of ``low-frequency" magnetization precession and non-periodic~\cite{VarelaManjarres2024,Freimuth2017} in the case of the ``high frequency'' demagnetization---that pumps  currents in the absence of any bias voltage~\footnote[3]{It is worth recalling arguments for  how terminology ``pumping'' has been attached~\cite{Freimuth2017} to 
		wider and wider classes of phenomena, while keeping them unified through common equations describing apparently vastly different situations, which we also follow here. Adiabatic quantum pumping was introduced by Thouless~\cite{Citro2023} as an effect, in which slow modulation of two or more
		external parameters of a quantum system, results in a net DC charge current in the absence of any externally applied bias voltage. In particular, Brouwer scattering-matrix-based formula~\cite{Brouwer1998} for such charge pumping from a quantum dot,  attached to two leads and modulated by two periodically changing gate voltages as fabricated  experimentally~\cite{Switkes1999}, was later applied~\cite{Tserkovnyak2005} to FM/NM bilayers to explain why precessing magnetization of FM layer generates pure (i.e., with no accompanied charge current) spin current flowing toward NM layer. In this approach, two time-periodic  components of magnetization (due to precession; the third component is fixed along the axis of precession)  replace two gate voltages of the original Brouwer scattering-matrix-based formula~\cite{Brouwer1998}.  Since the same formula explains both phenomena, such spin current generation is legitimately  also termed~\cite{Tserkovnyak2005} ``spin pumping.'' The same effect can be described by more general Keldysh GFs-based 
		expressions~\cite{Mahfouzi2012,VarelaManjarres2023,Dolui2020b}, which can also handle situations when magnetization changes non-periodically~\cite{Petrovic2018,Petrovic2021,Abbout2018} or is changing length~\cite{VarelaManjarres2024} instead of precessing. In the latter case, identical system---such as an illustrative example~\cite{Chen2009} of a single site with magnetization vector within one-dimensional tigh-binding chain---treated by the same time-dependent Keldysh GFs-based calculations will produce spin current~\cite{Chen2009,Petrovic2018} if magnetization vector is precessing  (which is identical~\cite{Chen2009} to the one~\cite{Tserkovnyak2005} computed from the Brouwer scattering-matrix-based  formula~\cite{Brouwer1998}), or it will produce  both~\cite{VarelaManjarres2024} spin and charge currents if its length is shrinking aperiodically~\cite{Freimuth2017} (as in demagnetization). This justifies usage of ``pumping''~\cite{Freimuth2017} of spin and charge terminology for all such effects.}.  Thus,  pumping of spin current by  demagnetization dynamics unveiled by Fig.~\ref{fig:fig2} replaces the need for phenomenological and ill-defined~\footnotemark[2] ``spin voltage'' as its driving mechanism~\cite{Buehlmann2020,Rouzegar2022,Seifert2023}.

	Findings of Fig.~\ref{fig:fig2} also provide microscopic  justification for previously conjectured  ``$dM/dt$ mechanism''~\cite{Choi2014,Lichtenberg2022,Rouzegar2022} from the fitting of experimental data---note that thereby motivated phenomenological theories require many qualitative assumptions, sometimes arriving at $I^z_\mathrm{S_\alpha} \propto \partial^2_t M^y$ (see Eq.~(47) in Ref.~\cite{Beens2022a}) which cannot be justified from our first-principles theory or experiments~\cite{Rouzegar2022}. 
	Once demagnetization sets in, the amplitude of generated spin current is primarily determined by the speed of demagnetization, which is larger in Co/Pt than in Co/W [compare values around \mbox{$\simeq 50$ fs} in Fig.~\ref{fig:fig2}(a) vs. Fig.~\ref{fig:fig2}(b)] due to difference in SOC-proximity effect~\cite{MarmolejoTejada2017,Dolui2020a,Zutic2019}. Via such proximity effect,  Pt or W layer can significantly  modify~\cite{MarmolejoTejada2017,Dolui2020a} properties of Co layer, of nanoscale thickness in experiments,  as also confirmed in ultrafast demagnetization experiments~\cite{Kuiper2014}. Experiments also find~\cite{Malinowski2008} that interlayer spin 
	current can affect demagnetization in self-consistent fashion---demagnetization acts as a primary nonequilibrium drive of spin current in THz frequency range in Fig.~\ref{fig:fig2}, which in turn further speeds up~\cite{Malinowski2008,Battiato2010} demagnetization as spin current carries away angular momentum from Co into good spin sinks~\cite{Tserkovnyak2005} like Pt or W.  The fact that  $I^z_{S_y} \neq 0$ is nonzero during  \mbox{$t < 25$ fs}, where    $\partial_t M^y \simeq 0$ [Fig.~\ref{fig:fig2}], can be explained by the presence~\cite{Suresh2023} of a second nonequilibrium drive, that is fsLP, which is initially the only one operative. Specific to Co, there can be a    temporal separation (\mbox{$\Delta t \sim 20$ fs})  between the commencement of spin
	current and demagnetization, as  noticed in some experiments~\cite{Shokeen2017}. In addition, since electrons respond instantaneously to fsLP, the start of oscillations in red curves in Fig.~\ref{fig:fig2} signals the true beginning of fsLP (see movies accompanying Ref.~\cite{Bajpai2019} for further illustration of how to properly identify the beginning of fsLP, which is not easy to  achieve by just visually  inspecting the gray curve in Fig.~\ref{fig:fig2}).
	
	We next {\em independently}  [i.e., without invoking any spin current and extraction of charge current from it via phenomenologically motivated  Eq.~\eqref{eq:ishe}] compute charge current 
	\begin{equation}\label{eq:chargecurrent}
		I_\mathrm{NM}^x(t) = \int\limits_\mathrm{NM} d^3\, r 
		j^x(\mathbf{r},t),
	\end{equation}
	flowing within NM=Pt or NM=W layer along the $x$-axis in Fig.~\ref{fig:fig1}, as expected from  ISHE~\cite{Saitoh2006} phenomenology.  Here $j^x(\mathbf{r},t)$ is the $x$-component of the vector of charge current density.  This procedure  allows us to test if intralayer charge current  satisfies   virtually 
	always   (for exceptions, see Ref.~\cite{Gorchon2022}) assumed~\cite{Seifert2016,Seifert2023,Wu2017,Zhang2017b,Rouzegar2022,Jechumtal2024,Liu2021} one-to-one relation [Eq.~\eqref{eq:ishe}] with interlayer spin current. Note that such relation has never been tested experimentally  because spin current is not~\cite{Saitoh2006,Tserkovnyak2005} directly measurable quantity and intralayer charge current is not directly accessible.   Since EM radiation in the FF region scanned experimentally is generated by the time-derivative of charge current [Eq.~\eqref{eq:eff}], rather than by current itself as sometimes {\em incorrectly}  assumed~\cite{Seifert2016,Rouzegar2022,Seifert2023,Jechumtal2024}, we plot in Fig.~\ref{fig:fig3}(a) time-derivative of spin current $\partial_t I^z_{S_y}$ in Co/Pt from Fig.~\ref{fig:fig2}(a) and compare it with time-derivative of charge current $\partial_t I_\mathrm{NM}^x$ within Pt in Fig.~\ref{fig:fig3}(b). The same comparison is performed in Fig.~\ref{fig:fig3}(d) vs. Fig.~\ref{fig:fig3}(e) for Co/W bilayer. As spin and charge current profiles differ, we conclude that Eq.~\eqref{eq:ishe} {\em is not justified} (thereby supporting the same conclusion of  experiments in Ref.~\cite{Gorchon2022}).
	
	Finally, we examine if charge current within the NM layer, generated by possibly multiple spin-to-charge conversion processes~\cite{Gorchon2022,Mahfouzi2014a,Agarwal2023} of incoming interlayer spin current in Fig.~\ref{fig:fig2}, is the sole contributor to outgoing THz radiation, as it has always been  assumed~\cite{Seifert2016,Wu2017,Zhang2017b,Rouzegar2022,Seifert2023,Jechumtal2024,Liu2021} when interpreting experiments. We note that despite being one of the main observables in experiments, both for single FM layer~\cite{Beaurepaire2004,Liu2021} and FM/NM bilayers~\cite{Seifert2016,Rouzegar2022,Seifert2023,Wu2017,Zhang2017b,Jechumtal2024,Liu2021}, EM radiation has been scarcely calculated. A  microscopic, as well as first principles, route for such an analysis has been formulated very recently as TDDFT+Jefimenko approach~\cite{Kefayati2024}. In it, charge current density  computed from TDDFT is  plugged into the Jefimenko formula~\cite{Jefimenko1966} for the electric field of EM radiation in the FF  region~\cite{McDonald1997}
	\begin{equation}\label{eq:eff}
		\mathbf{E}_\mathrm{FF}(\mathbf{r},t) =   \frac{1}{4\pi \varepsilon_0} \int\limits_{\substack{\mathrm{FM, \ or \  NM,} \\ \mathrm{or~FM/NM}}} \!\! d^3r' \, \biggl( \textbf{R} \frac{\partial_t \mathbf{j}_\mathrm{ret} \cdot \mathbf{R}}{c^2R^3} - \frac{\partial_t \textbf{j}_\mathrm{ret}}{c^2R} \biggl). 
	\end{equation}
	Here charge current density $\textbf{j}_\mathrm{ret}(\textbf{r},t) = \textbf{j}(\textbf{r}, t-R/c)$ is  evaluated at the retarded time $t-R/c$; $c$ is the velocity of light; $\mathbf{R}=\mathbf{r}-\mathbf{r'}$  is the vector from source at a point $\mathbf{r}'$ within FM/NM bilayer to the observation point $\mathbf{r}$; and  we also use shorthand notation  $R=|\mathbf{R}|$. The observation  point is chosen as $\mathbf{r}=10000 a_B\mathbf{e}_z$, where $a_B$ is the Bohr radius, and the origin of the coordinate system is in the lower left corner of Co layer [Fig.~\ref{fig:fig1}].
	The FF region is defined as the region where EM radiation decays as $\sim 1/R$, which isolates~\cite{McDonald1997} two terms from the full Jefimenko formula~\cite{Jefimenko1966,Griffiths1991} for electric field at arbitrary distance. Such full Jefimenko formulas for electric and magnetic field  of EM radiation can be viewed~\cite{Griffiths1991} as the proper time-dependent and time-retarded [via usage of  $\textbf{j}_\mathrm{ret}(\textbf{r},t)$] generalizations of the Coulomb and Biot-Savart laws, respectively. They are an integral solution of the Maxwell equations in the following approximations: fields vanish at infinity; their sources are confined to a finite region of space; and self-consistent effects, such as emitted EM radiation exerting backaction~\cite{Tancogne-Dejean2020,Philip2018} onto the source, can be neglected. 
	
	Figure~\ref{fig:fig3}(c) shows that the real-time signal of THz radiation contains contributions from charge currents within {\em both} Pt and Co layers, which can be spatially resolved by limits of integration in Eq.~\eqref{eq:eff}. The two contributions do not simply enhance each other due to their different phase. 
	The latter current, $I^x_\mathrm{FM}$, has been predicted~\cite{Kefayati2024,VarelaManjarres2024,Freimuth2017} as an additional consequence 
	of $\partial_t M^y(t)$ that acts as a nonequilibrium drive for electrons. Thus, time-dependence of $\partial_t M^y(t)$  always {\em concurrently} pumps {\em both} interlayer spin, $I^z_{S_\alpha}$, and intralayer charge, $I^x_\mathrm{FM}$  currents denoted in Fig.~\ref{fig:fig1}. This feature can be  contrasted to current pumping by precessing magnetization driven by low frequency EM radiation,  which typically generates only spin current~\cite{Tserkovnyak2005} (in the absence of SOC~\cite{Mahfouzi2012,VarelaManjarres2023}). Note that real-time THz signals in Figs.~\ref{fig:fig3}(c) and ~\ref{fig:fig3}(d) has maximum at the same time when fsLP (gray curve in the inset) reaches maximum. However, it also looks as if THz signals exists before fsLP. This is an illusion due to difficulty in showing  small oscillations that mark the true  beginning of fsLP, even if a much wider ordinate scale is used  than employed in Fig.~\ref{fig:fig3}(c) [one can also understand this issues by watching movies accompanying Ref.~\cite{Bajpai2019} which show how electrons start to flow even though fsLP is apparently still zero].
	
	When switching from Co/Pt to Co/W bilayer, the contribution to THz signal from spin-to-charge converted current flowing within W layer is, surprisingly, insignificant [blue dotted line in Fig.~\ref{fig:fig3}(f)] when compared to the contribution from charge current flowing within Co layer [green dashed line in Fig.~\ref{fig:fig3}(f)]. The same information, as charge current and THz signal are related via Jefimenko Eq.~\eqref{eq:eff}, is confirmed by comparing Fig.~\ref{fig:fig3}(e) vs. Fig.~\ref{fig:fig3}(b) where $\partial_t I^x_\mathrm{W}$ in the former case is an order of magnitude smaller than $\partial_t I^x_\mathrm{Pt}$ in the latter case.  Thus, this finding provides quite a different explanation for the observed change of sign of THz signal when switching from NM=Pt to NM=W layer---which is due to different proximitization~\cite{MarmolejoTejada2017,Dolui2020a} of Co by W---that is otherwise   attributed~\cite{Seifert2023} to the opposite sign of $\theta_\mathrm{SH}$ of Pt and W that subsequently inverts the charge current within NM layer assuming Eq.~\eqref{eq:ishe} holds true.  The fact that the charge current within the W layer is so small, thereby having very little effect on the total THz signal in Fig.~\ref{fig:fig3}(f), highlights the intricacies of both spin transport across FM/NM interfaces and ultrafast dynamics of electrons in general.  Regarding the latter issue, and unrelated to spin transport only, usage of transition metals as NM layers and ultrafast nature of electron dynamics can make possible highly nontrivial phenomena, such as  localization~\cite{Volkov2019} of excited electrons onto  the outer unfilled $d$-orbitals of W as a signature of transition from initially independent into dynamically correlated electron dynamics. In fact, precisely this phenomenon has been confirmed~\cite{Vos2023} for W very recently via  attosecond transient absorption spectroscopy~\cite{Volkov2019}. Such dynamical correlation effects can be isolated~\cite{Mrudul2024}  within the TDDFT framework by comparing calculations with frozen in time $v_H(\mathbf{r},t=0)$ and $v_{XC}(\mathbf{r},t=0)$ potentials in Eq.~\eqref{eq:tddft}, denoted 
	as ``independent particle approach'' in TDDFT as electrons are excited across energy levels of ground-state (GS) band structure irrespective~\cite{Adamantopoulos2022} 
	of how other electrons are time evolved,  vs. calculations where  these potentials are time-evolved to affect the wavefunction $\psi_j(\mathbf{r},t)$ in Eq.~\eqref{eq:tddft}. 
	
	{\em Models and Methods.}---The TDDFT calculations were performed via  extended-by-us~\cite{Kefayati2024,elkjefimenko} ELK package~\cite{Dewhurst2016,elk}. The thickness of Co is three, and of Pt or W is four, monolayers (MLs) along [100] crystallographic direction. TDDFT operates~\cite{Ullrich2011} with time-dependent version of  Kohn-Sham (KS) equation given by (using $\hbar=1$) 
	\begin{eqnarray}\label{eq:tddft}
		i\frac{\partial {\bm \psi}_j(\textbf{r},t)}{\partial t} & = & \bigg [ \frac{1}{2m_e} \bigg(-i\nabla+\frac{1}{c}\mathbf{A}_\mathrm{ext}(t) \bigg)^2 + v_s(\textbf{r},t) \nonumber\\
		& +& \frac{1}{2c} {\bm \sigma} \cdot \textbf{B}_s(\textbf{r},t)\nonumber\\
		& + &\frac{1}{4c^2}{\bm \sigma} \cdot \big( \nabla v_s(\textbf{r},t)\times -i\nabla\big) \bigg]\psi_j(\textbf{r},t),
	\end{eqnarray}
	where ${\bm \psi}_j(\textbf{r},t)$ are two-component Pauli spinors of KS quasiparticle; $m_e$ is the electron mass; $\mathbf{A}_\mathrm{ext}(t)$ is the vector potential of the applied fsLP; $v_s(\textbf{r},t) = v_{\mathrm{ext}}(\textbf{r},t) + v_{\mathrm{H}}(\textbf{r},t) + v_{\mathrm{XC}}(\textbf{r},t) $ is the effective KS potential, as the sum of the external potential $v_{\mathrm{ext}}$ provided by the nuclei (treated as point particles), the Hartree potential $v_{\mathrm{H}}$ and exchange-correlation  (XC) potential $v_\mathrm{XC}$;   $\mathbf{B}_s(\textbf{r},t)=\mathbf{B}_\mathrm{ext}(t)+\mathbf{B}_\mathrm{XC}(t)$ is the KS magnetic field with $\mathbf{B}_\mathrm{ext}$ being the external magnetic field and $\mathbf{B}_\mathrm{XC}$ the XC magnetic field; \mbox{${\bm \sigma}=(\sigma_x, \sigma_y, \sigma_z)$} is the vector of the Pauli matrices; and the last term on the RHS describes SOC which necessitates usage of noncollinear XC functionals~\cite{Eich2013a,Egidi2017} even when long-range noncollinearity of local magnetization does not play a significant role. The particle density of an interacting electronic system, as the fundamental quantity in (TD)DFT, is obtained as \mbox{$n(\mathbf{r},t)=\sum_j \psi_j^\dagger(\mathbf{r},t) \psi_j(\mathbf{r},t)$}. Similarly, magnetization density, as an additional fundamental quantity in noncollinear (TD)DFT, is obtained from \mbox{$\mathbf{m}(\mathbf{r},t)=\sum_j \psi_j^\dagger(\mathbf{r},t) {\bm \sigma} \psi_j(\mathbf{r},t)$}, so that total magnetization is given by \mbox{$\mathbf{M}(t)=\int\!\! d^3 r\, \mathbf{m}(\mathbf{r},t)$}. We employ adiabatic local density approximation (ALDA) for XC functional~\cite{Lacombe2023} within full-potential linearized augmented plane-wave method as implemented in the ELK code~\cite{Dewhurst2016,elk}. The GS is also obtained from ELK using noncollinear static DFT calculations with  LDA XC functional. The grid of $\mathbf{k}$ vectors is chosen as $7 \times 7$ for both Co/Pt or Co/W. After obtaining the GS, the dynamics for TDDFT calculations is generated by applying a Gaussian fsLP with: central wavelength \mbox{$800$ nm};  $\simeq 50$ fsLP  duration; the peak intensity of  $2.42$ TW/cm$^2$; and 22 mJ/cm$^2$ fluence. Since the wavelength of applied laser light is much larger than the supercell, we assume dipole approximation and disregard spatial dependence of $\mathbf{A}_\mathrm{ext}(t)$. 
	
	{\em Conclusions and Outlook}.---Using recently developed TDDFT+Jefimenko first-principles approach~\cite{Kefayati2024}, we  {\em independently}  compute  spin and charge currents in response to fsLP driving of Co/Pt and Co/W bilayers to find [Fig.~\ref{fig:fig3}] that in both cases \begin{equation}\label{eq:notequal}
		I^x_\mathrm{NM} (t) \neq \theta_\mathrm{SH} I_{S_y}^z(t). 
	\end{equation}
	This is in contrast to identity of their temporal  profiles  [Eq.~\eqref{eq:ishe}] that is virtually always assumed~\cite{Seifert2016,Seifert2023,Wu2017,Zhang2017b,Rouzegar2022,Jechumtal2024,Liu2021} when interpreting experiments, and also employed~\cite{Jechumtal2024} to extract $I_{S_y}^z(t)$ time profile  from measured  THz signals. However,  Eq.~\eqref{eq:ishe} can hardly be tested experimentally because spin current cannot be directly measured in general~\cite{Saitoh2006}, and measuring ultrafast charge current is outside the capabilities of presently available electronics. There are many  possible reasons~\cite{Gorchon2022,Schmidt2023} for breakdown  of one-to-one  relation between temporal profiles of $I_{S_y}^z(t)$ and $I^x_\mathrm{NM}(t)$, such as: multiple spin-to-charge conversion processes, at FM/NM interface or within NM bulk~\cite{Mahfouzi2014a}; spin memory loss~\cite{Belashchenko2016,Dolui2017,Gupta2020,Amin2016,Amin2016a}; reduced spin transmission~\cite{Wahada2022} across FM/NM interface; and localization~\cite{Vos2023,Volkov2019} of initially free excited electrons onto outer unfilled $d$-orbitals of transitions metals used as NM layer in the course of their ultrafast dynamics. We indeed find localization mechanism to be operative in the case of Co/W layer, as also confirmed in experiments on W alone~\cite{Vos2023}, which leads to {\em maximal violation} of virtually always assumed Eq.~\eqref{eq:ishe} as the charge current $I^x_W(t)$ [Fig.~\ref{fig:fig3}(e)] and its contribution to THz radiation [Fig.~\ref{fig:fig3}(f)] are minuscule. Possibility of such dramatic violation 
	of Eq.~\eqref{eq:ishe} is not surprising as relations like Eq.~\eqref{eq:ishe} stem from phenomenological assumption that one can simply translate concepts from linear-response (i.e., driven by bias voltage much smaller than the Fermi energy) steady-state transport phenomena~\cite{Sinova2015,VanTuan2016} to ultrafast-light-driven materials. On the other hand, using light of {800 nm} wavelength is equivalent to applying bias voltage of \mbox{$\simeq 1.55$ eV} (which some calculations on spintronic THz emitters even attempt to mimic by using steady-state quantum transport formulas at such finite bias voltage~\cite{Wahada2022}). On the top of it, one has to deal with ultrafast time evolution of electrons, which can bring a host of  nontrivial dynamical correlations effects~\cite{Mrudul2024} that {\em cannot} be conjectured by invoking a physical picture~\cite{Adamantopoulos2022}  where excited electrons simply transition across 
	bands calculated in the GS.
	
	Our first-principles analysis also reveals that \mbox{$I_{S_y}^z(t) \propto \partial_t M^y(t)$} [Fig.~\ref{fig:fig2}], thereby explaining that the origin of interlayer spin current is its pumping by time dependence of $M^y(t)$. Thus, the findings of Fig.~\ref{fig:fig2} displace the need for phenomenological ``spin voltage'' as the driving mechanism~\cite{Buehlmann2020,Rouzegar2022,Seifert2023} of interlayer spin current and its consequences [such as spin current $\propto \partial_t^2 M^y$~\cite{Beens2022a}, or $\propto \partial_t M^y$~\cite{Rouzegar2022} but for $M^y(t)$ measured on FM layer alone] that anyhow {\em cannot} be justified from microscopic quantum statistical framework~\footnotemark[2]. Finally, as already discussed in Refs.~\cite{Kefayati2024,VarelaManjarres2024}, the same time-dependence of $M^y(t)$ pumps {\em additional} charge current  within FM layer whose real-time THz emission signal is unravelled here [Fig.~\ref{fig:fig3}(c),(f)]. Such signal not been taken into account when interpreting experiments on THz emission from FM/NM bilayer~\cite{Seifert2016,Seifert2023,Wu2017,Zhang2017b,Rouzegar2022,Jechumtal2024,Liu2021} or single FM layer~\cite{Liu2021,Chu2024,Zhang2017b}. In the case of Co/W, $I^x_W(t)$ [Fig.~\ref{fig:fig3}(e)] and its radiation   [Fig.~\ref{fig:fig3}(f)] are  suppressed  by localization effects~\cite{Vos2023}.  In the case of Co/Pt, THz signal radiated by $I^x_\mathrm{Pt}(t)$ is present, but shifted in phase [Fig.~\ref{fig:fig3}(c)] with respect to signal radiated by $I^x_\mathrm{Co}(t)$, so that 
	two contributions dominate the total THz signal within {\em different}  time frames [Fig.~\ref{fig:fig3}(c)].

	\begin{acknowledgments}
		This research was supported by the US National Science Foundation (NSF) through the  University of Delaware Materials Research Science and Engineering Center, DMR-2011824. The supercomputing time was provided by DARWIN (Delaware Advanced Research Workforce and Innovation Network), which is supported by NSF Grant No. MRI-1919839.
	\end{acknowledgments}


\begin{thebibliography}{90}%
	\makeatletter
	\providecommand \@ifxundefined [1]{%
		\@ifx{#1\undefined}
	}%
	\providecommand \@ifnum [1]{%
		\ifnum #1\expandafter \@firstoftwo
		\else \expandafter \@secondoftwo
		\fi
	}%
	\providecommand \@ifx [1]{%
		\ifx #1\expandafter \@firstoftwo
		\else \expandafter \@secondoftwo
		\fi
	}%
	\providecommand \natexlab [1]{#1}%
	\providecommand \enquote  [1]{``#1''}%
	\providecommand \bibnamefont  [1]{#1}%
	\providecommand \bibfnamefont [1]{#1}%
	\providecommand \citenamefont [1]{#1}%
	\providecommand \href@noop [0]{\@secondoftwo}%
	\providecommand \href [0]{\begingroup \@sanitize@url \@href}%
	\providecommand \@href[1]{\@@startlink{#1}\@@href}%
	\providecommand \@@href[1]{\endgroup#1\@@endlink}%
	\providecommand \@sanitize@url [0]{\catcode `\\12\catcode `\$12\catcode `\&12\catcode `\#12\catcode `\^12\catcode `\_12\catcode `\%12\relax}%
	\providecommand \@@startlink[1]{}%
	\providecommand \@@endlink[0]{}%
	\providecommand \url  [0]{\begingroup\@sanitize@url \@url }%
	\providecommand \@url [1]{\endgroup\@href {#1}{\urlprefix }}%
	\providecommand \urlprefix  [0]{URL }%
	\providecommand \Eprint [0]{\href }%
	\providecommand \doibase [0]{https://doi.org/}%
	\providecommand \selectlanguage [0]{\@gobble}%
	\providecommand \bibinfo  [0]{\@secondoftwo}%
	\providecommand \bibfield  [0]{\@secondoftwo}%
	\providecommand \translation [1]{[#1]}%
	\providecommand \BibitemOpen [0]{}%
	\providecommand \bibitemStop [0]{}%
	\providecommand \bibitemNoStop [0]{.\EOS\space}%
	\providecommand \EOS [0]{\spacefactor3000\relax}%
	\providecommand \BibitemShut  [1]{\csname bibitem#1\endcsname}%
	\let\auto@bib@innerbib\@empty
	\bibitem [{\citenamefont {Seifert}\ \emph {et~al.}(2016)\citenamefont {Seifert}, \citenamefont {Jaiswal}, \citenamefont {Martens}, \citenamefont {Hannegan}, \citenamefont {Braun}, \citenamefont {Maldonado}, \citenamefont {Freimuth}, \citenamefont {Kronenberg}, \citenamefont {Henrizi}, \citenamefont {Radu} \emph {et~al.}}]{Seifert2016}%
	\BibitemOpen
	\bibfield  {author} {\bibinfo {author} {\bibfnamefont {T.}~\bibnamefont {Seifert}}, \bibinfo {author} {\bibfnamefont {S.}~\bibnamefont {Jaiswal}}, \bibinfo {author} {\bibfnamefont {U.}~\bibnamefont {Martens}}, \bibinfo {author} {\bibfnamefont {J.}~\bibnamefont {Hannegan}}, \bibinfo {author} {\bibfnamefont {L.}~\bibnamefont {Braun}}, \bibinfo {author} {\bibfnamefont {P.}~\bibnamefont {Maldonado}}, \bibinfo {author} {\bibfnamefont {F.}~\bibnamefont {Freimuth}}, \bibinfo {author} {\bibfnamefont {A.}~\bibnamefont {Kronenberg}}, \bibinfo {author} {\bibfnamefont {J.}~\bibnamefont {Henrizi}}, \bibinfo {author} {\bibfnamefont {I.}~\bibnamefont {Radu}}, \emph {et~al.},\ }\bibfield  {title} {\bibinfo {title} {Efficient metallic spintronic emitters of ultrabroadband terahertz radiation},\ }\href {https://doi.org/10.1038/nphoton.2016.91} {\bibfield  {journal} {\bibinfo  {journal} {Nat. Photonics}\ }\textbf {\bibinfo {volume} {10}},\ \bibinfo {pages} {483} (\bibinfo {year} {2016})}\BibitemShut {NoStop}%
	\bibitem [{\citenamefont {Seifert}\ \emph {et~al.}(2023)\citenamefont {Seifert}, \citenamefont {Go}, \citenamefont {Hayashi}, \citenamefont {Rouzegar}, \citenamefont {Freimuth}, \citenamefont {Ando}, \citenamefont {Mokrousov},\ and\ \citenamefont {Kampfrath}}]{Seifert2023}%
	\BibitemOpen
	\bibfield  {author} {\bibinfo {author} {\bibfnamefont {T.~S.}\ \bibnamefont {Seifert}}, \bibinfo {author} {\bibfnamefont {D.}~\bibnamefont {Go}}, \bibinfo {author} {\bibfnamefont {H.}~\bibnamefont {Hayashi}}, \bibinfo {author} {\bibfnamefont {R.}~\bibnamefont {Rouzegar}}, \bibinfo {author} {\bibfnamefont {F.}~\bibnamefont {Freimuth}}, \bibinfo {author} {\bibfnamefont {K.}~\bibnamefont {Ando}}, \bibinfo {author} {\bibfnamefont {Y.}~\bibnamefont {Mokrousov}},\ and\ \bibinfo {author} {\bibfnamefont {T.}~\bibnamefont {Kampfrath}},\ }\bibfield  {title} {\bibinfo {title} {Time-domain observation of ballistic orbital-angular-momentum currents with giant relaxation length in tungsten},\ }\href {https://doi.org/10.1038/s41565-023-01470-8} {\bibfield  {journal} {\bibinfo  {journal} {Nat. Nanotechnol.}\ }\textbf {\bibinfo {volume} {18}},\ \bibinfo {pages} {1132} (\bibinfo {year} {2023})}\BibitemShut {NoStop}%
	\bibitem [{\citenamefont {Wu}\ \emph {et~al.}(2017)\citenamefont {Wu}, \citenamefont {Elyasi}, \citenamefont {Qiu}, \citenamefont {Chen}, \citenamefont {Liu}, \citenamefont {Ke},\ and\ \citenamefont {Yang}}]{Wu2017}%
	\BibitemOpen
	\bibfield  {author} {\bibinfo {author} {\bibfnamefont {Y.}~\bibnamefont {Wu}}, \bibinfo {author} {\bibfnamefont {M.}~\bibnamefont {Elyasi}}, \bibinfo {author} {\bibfnamefont {X.}~\bibnamefont {Qiu}}, \bibinfo {author} {\bibfnamefont {M.}~\bibnamefont {Chen}}, \bibinfo {author} {\bibfnamefont {Y.}~\bibnamefont {Liu}}, \bibinfo {author} {\bibfnamefont {L.}~\bibnamefont {Ke}},\ and\ \bibinfo {author} {\bibfnamefont {H.}~\bibnamefont {Yang}},\ }\bibfield  {title} {\bibinfo {title} {High-performance {THz} emitters based on ferromagnetic/nonmagnetic heterostructures},\ }\href {https://doi.org/10.1002/adma.201603031} {\bibfield  {journal} {\bibinfo  {journal} {Adv. Mater.}\ }\textbf {\bibinfo {volume} {29}},\ \bibinfo {pages} {1603031} (\bibinfo {year} {2017})}\BibitemShut {NoStop}%
	\bibitem [{\citenamefont {Zhang}\ \emph {et~al.}(2017)\citenamefont {Zhang}, \citenamefont {Jin}, \citenamefont {Zhu}, \citenamefont {Zhu}, \citenamefont {Zhang}, \citenamefont {Ma},\ and\ \citenamefont {Yao}}]{Zhang2017b}%
	\BibitemOpen
	\bibfield  {author} {\bibinfo {author} {\bibfnamefont {S.}~\bibnamefont {Zhang}}, \bibinfo {author} {\bibfnamefont {Z.}~\bibnamefont {Jin}}, \bibinfo {author} {\bibfnamefont {Z.}~\bibnamefont {Zhu}}, \bibinfo {author} {\bibfnamefont {W.}~\bibnamefont {Zhu}}, \bibinfo {author} {\bibfnamefont {Z.}~\bibnamefont {Zhang}}, \bibinfo {author} {\bibfnamefont {G.}~\bibnamefont {Ma}},\ and\ \bibinfo {author} {\bibfnamefont {J.}~\bibnamefont {Yao}},\ }\bibfield  {title} {\bibinfo {title} {Bursts of efficient terahertz radiation with saturation effect from metal-based ferromagnetic heterostructures},\ }\href {https://doi.org/10.1088/1361-6463/aa9e43} {\bibfield  {journal} {\bibinfo  {journal} {J. Phys. D: Appl. Phys.}\ }\textbf {\bibinfo {volume} {51}},\ \bibinfo {pages} {034001} (\bibinfo {year} {2017})}\BibitemShut {NoStop}%
	\bibitem [{\citenamefont {Rouzegar}\ \emph {et~al.}(2022)\citenamefont {Rouzegar}, \citenamefont {Brandt}, \citenamefont {N\'advorn\'{\i}k}, \citenamefont {Reiss}, \citenamefont {Chekhov}, \citenamefont {Gueckstock}, \citenamefont {In}, \citenamefont {Wolf}, \citenamefont {Seifert}, \citenamefont {Brouwer} \emph {et~al.}}]{Rouzegar2022}%
	\BibitemOpen
	\bibfield  {author} {\bibinfo {author} {\bibfnamefont {R.}~\bibnamefont {Rouzegar}}, \bibinfo {author} {\bibfnamefont {L.}~\bibnamefont {Brandt}}, \bibinfo {author} {\bibfnamefont {L.~c.~v.}\ \bibnamefont {N\'advorn\'{\i}k}}, \bibinfo {author} {\bibfnamefont {D.~A.}\ \bibnamefont {Reiss}}, \bibinfo {author} {\bibfnamefont {A.~L.}\ \bibnamefont {Chekhov}}, \bibinfo {author} {\bibfnamefont {O.}~\bibnamefont {Gueckstock}}, \bibinfo {author} {\bibfnamefont {C.}~\bibnamefont {In}}, \bibinfo {author} {\bibfnamefont {M.}~\bibnamefont {Wolf}}, \bibinfo {author} {\bibfnamefont {T.~S.}\ \bibnamefont {Seifert}}, \bibinfo {author} {\bibfnamefont {P.~W.}\ \bibnamefont {Brouwer}}, \emph {et~al.},\ }\bibfield  {title} {\bibinfo {title} {Laser-induced terahertz spin transport in magnetic nanostructures arises from the same force as ultrafast demagnetization},\ }\href {https://doi.org/10.1103/PhysRevB.106.144427} {\bibfield  {journal} {\bibinfo  {journal} {Phys. Rev. B}\ }\textbf {\bibinfo {volume} {106}},\ \bibinfo {pages}
		{144427} (\bibinfo {year} {2022})}\BibitemShut {NoStop}%
	\bibitem [{\citenamefont {Jechumt\'al}\ \emph {et~al.}(2024)\citenamefont {Jechumt\'al}, \citenamefont {Rouzegar}, \citenamefont {Gueckstock}, \citenamefont {Denker}, \citenamefont {Hoppe}, \citenamefont {Remy}, \citenamefont {Seifert}, \citenamefont {Kuba\ifmmode \check{s}\else \v{s}\fi{}\ifmmode~\check{c}\else \v{c}\fi{}\'{\i}k}, \citenamefont {Woltersdorf}, \citenamefont {Brouwer}, \citenamefont {M\"unzenberg}, \citenamefont {Kampfrath},\ and\ \citenamefont {N\'advorn\'{\i}k}}]{Jechumtal2024}%
	\BibitemOpen
	\bibfield  {author} {\bibinfo {author} {\bibfnamefont {J.}~\bibnamefont {Jechumt\'al}}, \bibinfo {author} {\bibfnamefont {R.}~\bibnamefont {Rouzegar}}, \bibinfo {author} {\bibfnamefont {O.}~\bibnamefont {Gueckstock}}, \bibinfo {author} {\bibfnamefont {C.}~\bibnamefont {Denker}}, \bibinfo {author} {\bibfnamefont {W.}~\bibnamefont {Hoppe}}, \bibinfo {author} {\bibfnamefont {Q.}~\bibnamefont {Remy}}, \bibinfo {author} {\bibfnamefont {T.~S.}\ \bibnamefont {Seifert}}, \bibinfo {author} {\bibfnamefont {P.}~\bibnamefont {Kuba\ifmmode \check{s}\else \v{s}\fi{}\ifmmode~\check{c}\else \v{c}\fi{}\'{\i}k}}, \bibinfo {author} {\bibfnamefont {G.}~\bibnamefont {Woltersdorf}}, \bibinfo {author} {\bibfnamefont {P.~W.}\ \bibnamefont {Brouwer}}, \bibinfo {author} {\bibfnamefont {M.}~\bibnamefont {M\"unzenberg}}, \bibinfo {author} {\bibfnamefont {T.}~\bibnamefont {Kampfrath}},\ and\ \bibinfo {author} {\bibfnamefont {L.}~\bibnamefont {N\'advorn\'{\i}k}},\ }\bibfield  {title} {\bibinfo {title} {Accessing ultrafast spin-transport
			dynamics in copper using broadband terahertz spectroscopy},\ }\href {https://doi.org/10.1103/PhysRevLett.132.226703} {\bibfield  {journal} {\bibinfo  {journal} {Phys. Rev. Lett.}\ }\textbf {\bibinfo {volume} {132}},\ \bibinfo {pages} {226703} (\bibinfo {year} {2024})}\BibitemShut {NoStop}%
	\bibitem [{\citenamefont {Liu}\ \emph {et~al.}(2021)\citenamefont {Liu}, \citenamefont {Cheng}, \citenamefont {Xu}, \citenamefont {Vallobra}, \citenamefont {Eimer}, \citenamefont {Zhang}, \citenamefont {Wu}, \citenamefont {Nie},\ and\ \citenamefont {Zhao}}]{Liu2021}%
	\BibitemOpen
	\bibfield  {author} {\bibinfo {author} {\bibfnamefont {Y.}~\bibnamefont {Liu}}, \bibinfo {author} {\bibfnamefont {H.}~\bibnamefont {Cheng}}, \bibinfo {author} {\bibfnamefont {Y.}~\bibnamefont {Xu}}, \bibinfo {author} {\bibfnamefont {P.}~\bibnamefont {Vallobra}}, \bibinfo {author} {\bibfnamefont {S.}~\bibnamefont {Eimer}}, \bibinfo {author} {\bibfnamefont {X.}~\bibnamefont {Zhang}}, \bibinfo {author} {\bibfnamefont {X.}~\bibnamefont {Wu}}, \bibinfo {author} {\bibfnamefont {T.}~\bibnamefont {Nie}},\ and\ \bibinfo {author} {\bibfnamefont {W.}~\bibnamefont {Zhao}},\ }\bibfield  {title} {\bibinfo {title} {Separation of emission mechanisms in spintronic terahertz emitters},\ }\href {https://doi.org/10.1103/PhysRevB.104.064419} {\bibfield  {journal} {\bibinfo  {journal} {Phys. Rev. B}\ }\textbf {\bibinfo {volume} {104}},\ \bibinfo {pages} {064419} (\bibinfo {year} {2021})}\BibitemShut {NoStop}%
	\bibitem [{\citenamefont {Beaurepaire}\ \emph {et~al.}(1996)\citenamefont {Beaurepaire}, \citenamefont {Merle}, \citenamefont {Daunois},\ and\ \citenamefont {Bigot}}]{Beaurepaire1996}%
	\BibitemOpen
	\bibfield  {author} {\bibinfo {author} {\bibfnamefont {E.}~\bibnamefont {Beaurepaire}}, \bibinfo {author} {\bibfnamefont {J.-C.}\ \bibnamefont {Merle}}, \bibinfo {author} {\bibfnamefont {A.}~\bibnamefont {Daunois}},\ and\ \bibinfo {author} {\bibfnamefont {J.-Y.}\ \bibnamefont {Bigot}},\ }\bibfield  {title} {\bibinfo {title} {Ultrafast spin dynamics in ferromagnetic nickel},\ }\href {https://doi.org/10.1103/PhysRevLett.76.4250} {\bibfield  {journal} {\bibinfo  {journal} {Phys. Rev. Lett.}\ }\textbf {\bibinfo {volume} {76}},\ \bibinfo {pages} {4250} (\bibinfo {year} {1996})}\BibitemShut {NoStop}%
	\bibitem [{\citenamefont {Kirilyuk}\ \emph {et~al.}(2010)\citenamefont {Kirilyuk}, \citenamefont {Kimel},\ and\ \citenamefont {Rasing}}]{Kirilyuk2010}%
	\BibitemOpen
	\bibfield  {author} {\bibinfo {author} {\bibfnamefont {A.}~\bibnamefont {Kirilyuk}}, \bibinfo {author} {\bibfnamefont {A.~V.}\ \bibnamefont {Kimel}},\ and\ \bibinfo {author} {\bibfnamefont {T.}~\bibnamefont {Rasing}},\ }\bibfield  {title} {\bibinfo {title} {Ultrafast optical manipulation of magnetic order},\ }\href {https://doi.org/10.1103/revmodphys.82.2731} {\bibfield  {journal} {\bibinfo  {journal} {Rev. Mod. Phys.}\ }\textbf {\bibinfo {volume} {82}},\ \bibinfo {pages} {2731} (\bibinfo {year} {2010})}\BibitemShut {NoStop}%
	\bibitem [{\citenamefont {Scheid}\ \emph {et~al.}(2022)\citenamefont {Scheid}, \citenamefont {Remy}, \citenamefont {Lebègue}, \citenamefont {Malinowski},\ and\ \citenamefont {Mangin}}]{Scheid2022}%
	\BibitemOpen
	\bibfield  {author} {\bibinfo {author} {\bibfnamefont {P.}~\bibnamefont {Scheid}}, \bibinfo {author} {\bibfnamefont {Q.}~\bibnamefont {Remy}}, \bibinfo {author} {\bibfnamefont {S.}~\bibnamefont {Lebègue}}, \bibinfo {author} {\bibfnamefont {G.}~\bibnamefont {Malinowski}},\ and\ \bibinfo {author} {\bibfnamefont {S.}~\bibnamefont {Mangin}},\ }\bibfield  {title} {\bibinfo {title} {Light induced ultrafast magnetization dynamics in metallic compounds},\ }\href {https://doi.org/10.1016/j.jmmm.2022.169596} {\bibfield  {journal} {\bibinfo  {journal} {J. Magn. Magn. Mater.}\ }\textbf {\bibinfo {volume} {560}},\ \bibinfo {pages} {169596} (\bibinfo {year} {2022})}\BibitemShut {NoStop}%
	\bibitem [{\citenamefont {Chen}\ \emph {et~al.}(2025)\citenamefont {Chen}, \citenamefont {Adam}, \citenamefont {B\"{u}rgler}, \citenamefont {Wang}, \citenamefont {Lu}, \citenamefont {Pan}, \citenamefont {Heidtfeld}, \citenamefont {Greb}, \citenamefont {Liu}, \citenamefont {Liu}, \citenamefont {Wang}, \citenamefont {Schneider},\ and\ \citenamefont {Cao}}]{Chen2025}%
	\BibitemOpen
	\bibfield  {author} {\bibinfo {author} {\bibfnamefont {X.}~\bibnamefont {Chen}}, \bibinfo {author} {\bibfnamefont {R.}~\bibnamefont {Adam}}, \bibinfo {author} {\bibfnamefont {D.~E.}\ \bibnamefont {B\"{u}rgler}}, \bibinfo {author} {\bibfnamefont {F.}~\bibnamefont {Wang}}, \bibinfo {author} {\bibfnamefont {Z.}~\bibnamefont {Lu}}, \bibinfo {author} {\bibfnamefont {L.}~\bibnamefont {Pan}}, \bibinfo {author} {\bibfnamefont {S.}~\bibnamefont {Heidtfeld}}, \bibinfo {author} {\bibfnamefont {C.}~\bibnamefont {Greb}}, \bibinfo {author} {\bibfnamefont {M.}~\bibnamefont {Liu}}, \bibinfo {author} {\bibfnamefont {Q.}~\bibnamefont {Liu}}, \bibinfo {author} {\bibfnamefont {J.}~\bibnamefont {Wang}}, \bibinfo {author} {\bibfnamefont {C.~M.}\ \bibnamefont {Schneider}},\ and\ \bibinfo {author} {\bibfnamefont {D.}~\bibnamefont {Cao}},\ }\bibfield  {title} {\bibinfo {title} {Ultrafast demagnetization in ferromagnetic materials: Origins and progress},\ }\href {https://doi.org/10.1016/j.physrep.2024.10.008} {\bibfield  {journal}
		{\bibinfo  {journal} {Phys. Rep.}\ }\textbf {\bibinfo {volume} {1102}},\ \bibinfo {pages} {1} (\bibinfo {year} {2025})}\BibitemShut {NoStop}%
	\bibitem [{\citenamefont {Saitoh}\ \emph {et~al.}(2006)\citenamefont {Saitoh}, \citenamefont {Ueda}, \citenamefont {Miyajima},\ and\ \citenamefont {Tatara}}]{Saitoh2006}%
	\BibitemOpen
	\bibfield  {author} {\bibinfo {author} {\bibfnamefont {E.}~\bibnamefont {Saitoh}}, \bibinfo {author} {\bibfnamefont {M.}~\bibnamefont {Ueda}}, \bibinfo {author} {\bibfnamefont {H.}~\bibnamefont {Miyajima}},\ and\ \bibinfo {author} {\bibfnamefont {G.}~\bibnamefont {Tatara}},\ }\bibfield  {title} {\bibinfo {title} {Conversion of spin current into charge current at room temperature: Inverse spin-{Hall} effect},\ }\href {https://doi.org/https://doi.org/10.1063/1.2199473} {\bibfield  {journal} {\bibinfo  {journal} {Appl. Phys. Lett.}\ }\textbf {\bibinfo {volume} {88}},\ \bibinfo {pages} {182509} (\bibinfo {year} {2006})}\BibitemShut {NoStop}%
	\bibitem [{\citenamefont {Jungfleisch}\ \emph {et~al.}(2018)\citenamefont {Jungfleisch}, \citenamefont {Zhang}, \citenamefont {Zhang}, \citenamefont {Pearson}, \citenamefont {Schaller}, \citenamefont {Wen},\ and\ \citenamefont {Hoffmann}}]{Jungfleisch2018a}%
	\BibitemOpen
	\bibfield  {author} {\bibinfo {author} {\bibfnamefont {M.~B.}\ \bibnamefont {Jungfleisch}}, \bibinfo {author} {\bibfnamefont {Q.}~\bibnamefont {Zhang}}, \bibinfo {author} {\bibfnamefont {W.}~\bibnamefont {Zhang}}, \bibinfo {author} {\bibfnamefont {J.~E.}\ \bibnamefont {Pearson}}, \bibinfo {author} {\bibfnamefont {R.~D.}\ \bibnamefont {Schaller}}, \bibinfo {author} {\bibfnamefont {H.}~\bibnamefont {Wen}},\ and\ \bibinfo {author} {\bibfnamefont {A.}~\bibnamefont {Hoffmann}},\ }\bibfield  {title} {\bibinfo {title} {Control of terahertz emission by ultrafast spin-charge current conversion at {Rashba} interfaces},\ }\href {https://doi.org/10.1103/PhysRevLett.120.207207} {\bibfield  {journal} {\bibinfo  {journal} {Phys. Rev. Lett.}\ }\textbf {\bibinfo {volume} {120}},\ \bibinfo {pages} {207207} (\bibinfo {year} {2018})}\BibitemShut {NoStop}%
	\bibitem [{\citenamefont {Gueckstock}\ \emph {et~al.}(2021)\citenamefont {Gueckstock}, \citenamefont {N{\'{a}}dvorn{\'{\i}}k}, \citenamefont {Gradhand}, \citenamefont {Seifert}, \citenamefont {Bierhance}, \citenamefont {Rouzegar}, \citenamefont {Wolf}, \citenamefont {Vafaee}, \citenamefont {Cramer}, \citenamefont {Syskaki} \emph {et~al.}}]{Gueckstock2021}%
	\BibitemOpen
	\bibfield  {author} {\bibinfo {author} {\bibfnamefont {O.}~\bibnamefont {Gueckstock}}, \bibinfo {author} {\bibfnamefont {L.}~\bibnamefont {N{\'{a}}dvorn{\'{\i}}k}}, \bibinfo {author} {\bibfnamefont {M.}~\bibnamefont {Gradhand}}, \bibinfo {author} {\bibfnamefont {T.~S.}\ \bibnamefont {Seifert}}, \bibinfo {author} {\bibfnamefont {G.}~\bibnamefont {Bierhance}}, \bibinfo {author} {\bibfnamefont {R.}~\bibnamefont {Rouzegar}}, \bibinfo {author} {\bibfnamefont {M.}~\bibnamefont {Wolf}}, \bibinfo {author} {\bibfnamefont {M.}~\bibnamefont {Vafaee}}, \bibinfo {author} {\bibfnamefont {J.}~\bibnamefont {Cramer}}, \bibinfo {author} {\bibfnamefont {M.~A.}\ \bibnamefont {Syskaki}}, \emph {et~al.},\ }\bibfield  {title} {\bibinfo {title} {Terahertz spin-to-charge conversion by interfacial skew scattering in metallic bilayers},\ }\href {https://doi.org/10.1002/adma.202006281} {\bibfield  {journal} {\bibinfo  {journal} {Adv. Mater.}\ }\textbf {\bibinfo {volume} {33}},\ \bibinfo {pages} {2006281} (\bibinfo {year}
		{2021})}\BibitemShut {NoStop}%
	\bibitem [{\citenamefont {Wang}\ \emph {et~al.}(2023)\citenamefont {Wang}, \citenamefont {Li}, \citenamefont {Cheng}, \citenamefont {Liu}, \citenamefont {Cui}, \citenamefont {Huang}, \citenamefont {Xiong}, \citenamefont {Yang}, \citenamefont {Huang}, \citenamefont {Wang} \emph {et~al.}}]{Wang2023}%
	\BibitemOpen
	\bibfield  {author} {\bibinfo {author} {\bibfnamefont {Y.}~\bibnamefont {Wang}}, \bibinfo {author} {\bibfnamefont {W.}~\bibnamefont {Li}}, \bibinfo {author} {\bibfnamefont {H.}~\bibnamefont {Cheng}}, \bibinfo {author} {\bibfnamefont {Z.}~\bibnamefont {Liu}}, \bibinfo {author} {\bibfnamefont {Z.}~\bibnamefont {Cui}}, \bibinfo {author} {\bibfnamefont {J.}~\bibnamefont {Huang}}, \bibinfo {author} {\bibfnamefont {B.}~\bibnamefont {Xiong}}, \bibinfo {author} {\bibfnamefont {J.}~\bibnamefont {Yang}}, \bibinfo {author} {\bibfnamefont {H.}~\bibnamefont {Huang}}, \bibinfo {author} {\bibfnamefont {J.}~\bibnamefont {Wang}}, \emph {et~al.},\ }\bibfield  {title} {\bibinfo {title} {Enhancement of spintronic terahertz emission enabled by increasing {Hall} angle and interfacial skew scattering},\ }\href {https://doi.org/10.1038/s42005-023-01402-x} {\bibfield  {journal} {\bibinfo  {journal} {Commun. Phys.}\ }\textbf {\bibinfo {volume} {6}},\ \bibinfo {pages} {280} (\bibinfo {year} {2023})}\BibitemShut {NoStop}%
	\bibitem [{\citenamefont {Lu}\ \emph {et~al.}(2020)\citenamefont {Lu}, \citenamefont {Zhao}, \citenamefont {Battiato}, \citenamefont {Wu},\ and\ \citenamefont {Yuan}}]{Lu2020}%
	\BibitemOpen
	\bibfield  {author} {\bibinfo {author} {\bibfnamefont {W.-T.}\ \bibnamefont {Lu}}, \bibinfo {author} {\bibfnamefont {Y.}~\bibnamefont {Zhao}}, \bibinfo {author} {\bibfnamefont {M.}~\bibnamefont {Battiato}}, \bibinfo {author} {\bibfnamefont {Y.}~\bibnamefont {Wu}},\ and\ \bibinfo {author} {\bibfnamefont {Z.}~\bibnamefont {Yuan}},\ }\bibfield  {title} {\bibinfo {title} {Interface reflectivity of a superdiffusive spin current in ultrafast demagnetization and terahertz emission},\ }\href {https://doi.org/10.1103/PhysRevB.101.014435} {\bibfield  {journal} {\bibinfo  {journal} {Phys. Rev. B}\ }\textbf {\bibinfo {volume} {101}},\ \bibinfo {pages} {014435} (\bibinfo {year} {2020})}\BibitemShut {NoStop}%
	\bibitem [{\citenamefont {Mahfouzi}\ \emph {et~al.}(2014)\citenamefont {Mahfouzi}, \citenamefont {Nagaosa},\ and\ \citenamefont {Nikoli\ifmmode~\acute{c}\else \'{c}\fi{}}}]{Mahfouzi2014a}%
	\BibitemOpen
	\bibfield  {author} {\bibinfo {author} {\bibfnamefont {F.}~\bibnamefont {Mahfouzi}}, \bibinfo {author} {\bibfnamefont {N.}~\bibnamefont {Nagaosa}},\ and\ \bibinfo {author} {\bibfnamefont {B.~K.}\ \bibnamefont {Nikoli\ifmmode~\acute{c}\else \'{c}\fi{}}},\ }\bibfield  {title} {\bibinfo {title} {Spin-to-charge conversion in lateral and vertical topological-insulator/ferromagnet heterostructures with microwave-driven precessing magnetization},\ }\href {https://doi.org/10.1103/PhysRevB.90.115432} {\bibfield  {journal} {\bibinfo  {journal} {Phys. Rev. B}\ }\textbf {\bibinfo {volume} {90}},\ \bibinfo {pages} {115432} (\bibinfo {year} {2014})}\BibitemShut {NoStop}%
	\bibitem [{Note1()}]{Note1}%
	\BibitemOpen
	\bibinfo {note} {Assuming that spin accumulation alone allows one to describe demagnetization and concomitant processes like spin current generation is equivalent to using only diagonal elements of time-dependent nonequilibrium density matrix. In contrast, recent construction~\cite {Mrudul2024} of such density matrix from TDDFT calculations shows significance of its off-diagonal elements (when represented in the basis of KS orbitals in the GS).}\BibitemShut {Stop}%
	\bibitem [{\citenamefont {Bühlmann}\ \emph {et~al.}(2020)\citenamefont {Bühlmann}, \citenamefont {Saerens}, \citenamefont {Vaterlaus},\ and\ \citenamefont {Acremann}}]{Buehlmann2020}%
	\BibitemOpen
	\bibfield  {author} {\bibinfo {author} {\bibfnamefont {K.}~\bibnamefont {Bühlmann}}, \bibinfo {author} {\bibfnamefont {G.}~\bibnamefont {Saerens}}, \bibinfo {author} {\bibfnamefont {A.}~\bibnamefont {Vaterlaus}},\ and\ \bibinfo {author} {\bibfnamefont {Y.}~\bibnamefont {Acremann}},\ }\bibfield  {title} {\bibinfo {title} {Detection of femtosecond spin voltage pulses in a thin iron film},\ }\href {https://doi.org/10.1063/4.0000037} {\bibfield  {journal} {\bibinfo  {journal} {Struct. Dyn.}\ }\textbf {\bibinfo {volume} {7}},\ \bibinfo {pages} {065101} (\bibinfo {year} {2020})}\BibitemShut {NoStop}%
	\bibitem [{Note2()}]{Note2}%
	\BibitemOpen
	\bibinfo {note} {Note that defining the distribution function of electrons in far-from-equilibrium quantum systems is required to extract meaningful effective spin-dependent chemical potentials from it and then define ``spin voltage'' as their difference. However, this is a daunting task from a rigorous viewpoint. This is because distribution function of equilibrium quantum statistical mechanics [where it is justified by diagonal form of density matrix in the basis of energy eigenstates, such as from grand canonical density matrix \protect \mbox {$\protect \hat {\rho }_\protect \mathrm {eq} = \DOTSB \sum@ \slimits@ _n f(E_n)|E_n\rangle \langle E_n|$} we read off standard {\protect \em thermal} Fermi-Dirac distribution $f(E_n)$] or semiclassical nonequilibrium theories (such as the Boltzmann equation, sometimes used to model ultrafast demagnetization while requiring a host of additional assumptions~\cite {Nenno2018,Nenno2019,Hurst2018}) is replaced in nonequilibrium quantum statistical mechanics with
		single-time-dependent density matrix or its generalization in the form of two-times-dependent lesser Keldysh Green's functions (GF)~\cite {Stefanucci2025}. Even when the distribution function is extracted via some ansatz (typically by assuming that Keldysh GFs satisfy equilibrium-like relations even in nonequilibrium), it typically has a {\protect \em nonthermal} form that cannot be described by a single effective chemical potential and single effective temperature~\cite {Kuenzel2024,Takei2019}.}\BibitemShut {Stop}%
	\bibitem [{\citenamefont {Battiato}\ \emph {et~al.}(2010)\citenamefont {Battiato}, \citenamefont {Carva},\ and\ \citenamefont {Oppeneer}}]{Battiato2010}%
	\BibitemOpen
	\bibfield  {author} {\bibinfo {author} {\bibfnamefont {M.}~\bibnamefont {Battiato}}, \bibinfo {author} {\bibfnamefont {K.}~\bibnamefont {Carva}},\ and\ \bibinfo {author} {\bibfnamefont {P.~M.}\ \bibnamefont {Oppeneer}},\ }\bibfield  {title} {\bibinfo {title} {Superdiffusive spin transport as a mechanism of ultrafast demagnetization},\ }\href {https://doi.org/10.1103/PhysRevLett.105.027203} {\bibfield  {journal} {\bibinfo  {journal} {Phys. Rev. Lett.}\ }\textbf {\bibinfo {volume} {105}},\ \bibinfo {pages} {027203} (\bibinfo {year} {2010})}\BibitemShut {NoStop}%
	\bibitem [{\citenamefont {Nenno}\ \emph {et~al.}(2018)\citenamefont {Nenno}, \citenamefont {Rethfeld},\ and\ \citenamefont {Schneider}}]{Nenno2018}%
	\BibitemOpen
	\bibfield  {author} {\bibinfo {author} {\bibfnamefont {D.~M.}\ \bibnamefont {Nenno}}, \bibinfo {author} {\bibfnamefont {B.}~\bibnamefont {Rethfeld}},\ and\ \bibinfo {author} {\bibfnamefont {H.~C.}\ \bibnamefont {Schneider}},\ }\bibfield  {title} {\bibinfo {title} {Particle-in-cell simulation of ultrafast hot-carrier transport in {Fe/Au} heterostructures},\ }\href {https://doi.org/10.1103/PhysRevB.98.224416} {\bibfield  {journal} {\bibinfo  {journal} {Phys. Rev. B}\ }\textbf {\bibinfo {volume} {98}},\ \bibinfo {pages} {224416} (\bibinfo {year} {2018})}\BibitemShut {NoStop}%
	\bibitem [{\citenamefont {Nenno}\ \emph {et~al.}(2019)\citenamefont {Nenno}, \citenamefont {Binder},\ and\ \citenamefont {Schneider}}]{Nenno2019}%
	\BibitemOpen
	\bibfield  {author} {\bibinfo {author} {\bibfnamefont {D.~M.}\ \bibnamefont {Nenno}}, \bibinfo {author} {\bibfnamefont {R.}~\bibnamefont {Binder}},\ and\ \bibinfo {author} {\bibfnamefont {H.~C.}\ \bibnamefont {Schneider}},\ }\bibfield  {title} {\bibinfo {title} {Simulation of hot-carrier dynamics and terahertz emission in laser-excited metallic bilayers},\ }\href {https://doi.org/10.1103/physrevapplied.11.054083} {\bibfield  {journal} {\bibinfo  {journal} {Phys. Rev. Appl.}\ }\textbf {\bibinfo {volume} {11}},\ \bibinfo {pages} {054083} (\bibinfo {year} {2019})}\BibitemShut {NoStop}%
	\bibitem [{\citenamefont {Hurst}\ \emph {et~al.}(2018)\citenamefont {Hurst}, \citenamefont {Hervieux},\ and\ \citenamefont {Manfredi}}]{Hurst2018}%
	\BibitemOpen
	\bibfield  {author} {\bibinfo {author} {\bibfnamefont {J.}~\bibnamefont {Hurst}}, \bibinfo {author} {\bibfnamefont {P.-A.}\ \bibnamefont {Hervieux}},\ and\ \bibinfo {author} {\bibfnamefont {G.}~\bibnamefont {Manfredi}},\ }\bibfield  {title} {\bibinfo {title} {Spin current generation by ultrafast laser pulses in ferromagnetic nickel films},\ }\href {https://doi.org/10.1103/PhysRevB.97.014424} {\bibfield  {journal} {\bibinfo  {journal} {Phys. Rev. B}\ }\textbf {\bibinfo {volume} {97}},\ \bibinfo {pages} {014424} (\bibinfo {year} {2018})}\BibitemShut {NoStop}%
	\bibitem [{\citenamefont {Beens}\ \emph {et~al.}(2022)\citenamefont {Beens}, \citenamefont {Duine},\ and\ \citenamefont {Koopmans}}]{Beens2022a}%
	\BibitemOpen
	\bibfield  {author} {\bibinfo {author} {\bibfnamefont {M.}~\bibnamefont {Beens}}, \bibinfo {author} {\bibfnamefont {R.~A.}\ \bibnamefont {Duine}},\ and\ \bibinfo {author} {\bibfnamefont {B.}~\bibnamefont {Koopmans}},\ }\bibfield  {title} {\bibinfo {title} {Modeling ultrafast demagnetization and spin transport: The interplay of spin-polarized electrons and thermal magnons},\ }\href {https://doi.org/10.1103/PhysRevB.105.144420} {\bibfield  {journal} {\bibinfo  {journal} {Phys. Rev. B}\ }\textbf {\bibinfo {volume} {105}},\ \bibinfo {pages} {144420} (\bibinfo {year} {2022})}\BibitemShut {NoStop}%
	\bibitem [{\citenamefont {K\"unzel}\ \emph {et~al.}(2024)\citenamefont {K\"unzel}, \citenamefont {Erpenbeck}, \citenamefont {Werner}, \citenamefont {Arrigoni}, \citenamefont {Gull}, \citenamefont {Cohen},\ and\ \citenamefont {Eckstein}}]{Kuenzel2024}%
	\BibitemOpen
	\bibfield  {author} {\bibinfo {author} {\bibfnamefont {F.}~\bibnamefont {K\"unzel}}, \bibinfo {author} {\bibfnamefont {A.}~\bibnamefont {Erpenbeck}}, \bibinfo {author} {\bibfnamefont {D.}~\bibnamefont {Werner}}, \bibinfo {author} {\bibfnamefont {E.}~\bibnamefont {Arrigoni}}, \bibinfo {author} {\bibfnamefont {E.}~\bibnamefont {Gull}}, \bibinfo {author} {\bibfnamefont {G.}~\bibnamefont {Cohen}},\ and\ \bibinfo {author} {\bibfnamefont {M.}~\bibnamefont {Eckstein}},\ }\bibfield  {title} {\bibinfo {title} {Numerically exact simulation of photodoped {Mott} insulators},\ }\href {https://doi.org/10.1103/PhysRevLett.132.176501} {\bibfield  {journal} {\bibinfo  {journal} {Phys. Rev. Lett.}\ }\textbf {\bibinfo {volume} {132}},\ \bibinfo {pages} {176501} (\bibinfo {year} {2024})}\BibitemShut {NoStop}%
	\bibitem [{\citenamefont {Takei}(2019)}]{Takei2019}%
	\BibitemOpen
	\bibfield  {author} {\bibinfo {author} {\bibfnamefont {S.}~\bibnamefont {Takei}},\ }\bibfield  {title} {\bibinfo {title} {Spin transport in an electrically driven magnon gas near {Bose-Einstein} condensation: {Hartree-Fock-Keldysh} theory},\ }\href {https://doi.org/10.1103/physrevb.100.134440} {\bibfield  {journal} {\bibinfo  {journal} {Phys. Rev. B}\ }\textbf {\bibinfo {volume} {100}},\ \bibinfo {pages} {134440} (\bibinfo {year} {2019})}\BibitemShut {NoStop}%
	\bibitem [{\citenamefont {Krieger}\ \emph {et~al.}(2015)\citenamefont {Krieger}, \citenamefont {Dewhurst}, \citenamefont {Elliott}, \citenamefont {Sharma},\ and\ \citenamefont {Gross}}]{Krieger2015}%
	\BibitemOpen
	\bibfield  {author} {\bibinfo {author} {\bibfnamefont {K.}~\bibnamefont {Krieger}}, \bibinfo {author} {\bibfnamefont {J.~K.}\ \bibnamefont {Dewhurst}}, \bibinfo {author} {\bibfnamefont {P.}~\bibnamefont {Elliott}}, \bibinfo {author} {\bibfnamefont {S.}~\bibnamefont {Sharma}},\ and\ \bibinfo {author} {\bibfnamefont {E.~K.~U.}\ \bibnamefont {Gross}},\ }\bibfield  {title} {\bibinfo {title} {Laser-induced demagnetization at ultrashort time scales: Predictions of {TDDFT}},\ }\href {https://doi.org/10.1021/acs.jctc.5b00621} {\bibfield  {journal} {\bibinfo  {journal} {J. Chem. Theory Comput.}\ }\textbf {\bibinfo {volume} {11}},\ \bibinfo {pages} {4870} (\bibinfo {year} {2015})}\BibitemShut {NoStop}%
	\bibitem [{\citenamefont {Zhang}\ and\ \citenamefont {H\"ubner}(2000)}]{Zhang2000a}%
	\BibitemOpen
	\bibfield  {author} {\bibinfo {author} {\bibfnamefont {G.~P.}\ \bibnamefont {Zhang}}\ and\ \bibinfo {author} {\bibfnamefont {W.}~\bibnamefont {H\"ubner}},\ }\bibfield  {title} {\bibinfo {title} {Laser-induced ultrafast demagnetization in ferromagnetic metals},\ }\href {https://doi.org/10.1103/PhysRevLett.85.3025} {\bibfield  {journal} {\bibinfo  {journal} {Phys. Rev. Lett.}\ }\textbf {\bibinfo {volume} {85}},\ \bibinfo {pages} {3025} (\bibinfo {year} {2000})}\BibitemShut {NoStop}%
	\bibitem [{\citenamefont {Sinova}\ \emph {et~al.}(2015)\citenamefont {Sinova}, \citenamefont {Valenzuela}, \citenamefont {Wunderlich}, \citenamefont {Back},\ and\ \citenamefont {Jungwirth}}]{Sinova2015}%
	\BibitemOpen
	\bibfield  {author} {\bibinfo {author} {\bibfnamefont {J.}~\bibnamefont {Sinova}}, \bibinfo {author} {\bibfnamefont {S.~O.}\ \bibnamefont {Valenzuela}}, \bibinfo {author} {\bibfnamefont {J.}~\bibnamefont {Wunderlich}}, \bibinfo {author} {\bibfnamefont {C.~H.}\ \bibnamefont {Back}},\ and\ \bibinfo {author} {\bibfnamefont {T.}~\bibnamefont {Jungwirth}},\ }\bibfield  {title} {\bibinfo {title} {Spin hall effects},\ }\href {https://doi.org/10.1103/RevModPhys.87.1213} {\bibfield  {journal} {\bibinfo  {journal} {Rev. Mod. Phys.}\ }\textbf {\bibinfo {volume} {87}},\ \bibinfo {pages} {1213} (\bibinfo {year} {2015})}\BibitemShut {NoStop}%
	\bibitem [{\citenamefont {Gorchon}\ \emph {et~al.}(2022)\citenamefont {Gorchon}, \citenamefont {Mangin}, \citenamefont {Hehn},\ and\ \citenamefont {Malinowski}}]{Gorchon2022}%
	\BibitemOpen
	\bibfield  {author} {\bibinfo {author} {\bibfnamefont {J.}~\bibnamefont {Gorchon}}, \bibinfo {author} {\bibfnamefont {S.}~\bibnamefont {Mangin}}, \bibinfo {author} {\bibfnamefont {M.}~\bibnamefont {Hehn}},\ and\ \bibinfo {author} {\bibfnamefont {G.}~\bibnamefont {Malinowski}},\ }\bibfield  {title} {\bibinfo {title} {Is terahertz emission a good probe of the spin current attenuation length?},\ }\href {https://doi.org/10.1063/5.0097448} {\bibfield  {journal} {\bibinfo  {journal} {Appl. Phys. Lett.}\ }\textbf {\bibinfo {volume} {121}},\ \bibinfo {pages} {012402} (\bibinfo {year} {2022})}\BibitemShut {NoStop}%
	\bibitem [{\citenamefont {Schmidt}\ \emph {et~al.}(2023)\citenamefont {Schmidt}, \citenamefont {Das-Mohapatra},\ and\ \citenamefont {Papaioannou}}]{Schmidt2023}%
	\BibitemOpen
	\bibfield  {author} {\bibinfo {author} {\bibfnamefont {G.}~\bibnamefont {Schmidt}}, \bibinfo {author} {\bibfnamefont {B.}~\bibnamefont {Das-Mohapatra}},\ and\ \bibinfo {author} {\bibfnamefont {E.~T.}\ \bibnamefont {Papaioannou}},\ }\bibfield  {title} {\bibinfo {title} {Charge dynamics in spintronic terahertz emitters},\ }\href {https://doi.org/10.1103/PhysRevApplied.19.L041001} {\bibfield  {journal} {\bibinfo  {journal} {Phys. Rev. Appl.}\ }\textbf {\bibinfo {volume} {19}},\ \bibinfo {pages} {L041001} (\bibinfo {year} {2023})}\BibitemShut {NoStop}%
	\bibitem [{\citenamefont {Varela-Manjarres}\ \emph {et~al.}(2024)\citenamefont {Varela-Manjarres}, \citenamefont {Kefayati}, \citenamefont {Jungfleisch}, \citenamefont {Xiao},\ and\ \citenamefont {Nikoli\ifmmode~\acute{c}\else \'{c}\fi{}}}]{VarelaManjarres2024}%
	\BibitemOpen
	\bibfield  {author} {\bibinfo {author} {\bibfnamefont {J.}~\bibnamefont {Varela-Manjarres}}, \bibinfo {author} {\bibfnamefont {A.}~\bibnamefont {Kefayati}}, \bibinfo {author} {\bibfnamefont {M.~B.}\ \bibnamefont {Jungfleisch}}, \bibinfo {author} {\bibfnamefont {J.~Q.}\ \bibnamefont {Xiao}},\ and\ \bibinfo {author} {\bibfnamefont {B.~K.}\ \bibnamefont {Nikoli\ifmmode~\acute{c}\else \'{c}\fi{}}},\ }\bibfield  {title} {\bibinfo {title} {Charge and spin current pumping by ultrafast demagnetization dynamics},\ }\href {https://doi.org/10.1103/PhysRevB.110.L060410} {\bibfield  {journal} {\bibinfo  {journal} {Phys. Rev. B}\ }\textbf {\bibinfo {volume} {110}},\ \bibinfo {pages} {L060410} (\bibinfo {year} {2024})}\BibitemShut {NoStop}%
	\bibitem [{\citenamefont {Freimuth}\ \emph {et~al.}(2017)\citenamefont {Freimuth}, \citenamefont {Bl\"ugel},\ and\ \citenamefont {Mokrousov}}]{Freimuth2017}%
	\BibitemOpen
	\bibfield  {author} {\bibinfo {author} {\bibfnamefont {F.}~\bibnamefont {Freimuth}}, \bibinfo {author} {\bibfnamefont {S.}~\bibnamefont {Bl\"ugel}},\ and\ \bibinfo {author} {\bibfnamefont {Y.}~\bibnamefont {Mokrousov}},\ }\bibfield  {title} {\bibinfo {title} {Charge pumping driven by the laser-induced dynamics of the exchange splitting},\ }\href {https://doi.org/10.1103/PhysRevB.95.094434} {\bibfield  {journal} {\bibinfo  {journal} {Phys. Rev. B}\ }\textbf {\bibinfo {volume} {95}},\ \bibinfo {pages} {094434} (\bibinfo {year} {2017})}\BibitemShut {NoStop}%
	\bibitem [{\citenamefont {Cohen}(2014)}]{Cohen2014}%
	\BibitemOpen
	\bibfield  {author} {\bibinfo {author} {\bibfnamefont {M.~X.}\ \bibnamefont {Cohen}},\ }\href {https://doi.org/10.7551/mitpress/9609.001.0001} {\emph {\bibinfo {title} {Analyzing Neural Time Series Data: Theory and Practice}}}\ (\bibinfo  {publisher} {The MIT Press, Cambridge,},\ \bibinfo {year} {2014})\BibitemShut {NoStop}%
	\bibitem [{exp()}]{explaintdf}%
	\BibitemOpen
	\href@noop {} {\bibinfo {title} {Note that time domain filtering [35] offers advantages of frequency-domain filtering when dealing with transient events as it provides more direct control over the signal shape, particularly when dealing with sharp changes or sudden variations in the signal that might be obscured in the frequency domain. {I}t also allows for easier implementation of certain types of filters that are difficult to design in the frequency domain}}\BibitemShut {NoStop}%
	\bibitem [{\citenamefont {Shokeen}\ \emph {et~al.}(2017)\citenamefont {Shokeen}, \citenamefont {Sanchez~Piaia}, \citenamefont {Bigot}, \citenamefont {M\"uller}, \citenamefont {Elliott}, \citenamefont {Dewhurst}, \citenamefont {Sharma},\ and\ \citenamefont {Gross}}]{Shokeen2017}%
	\BibitemOpen
	\bibfield  {author} {\bibinfo {author} {\bibfnamefont {V.}~\bibnamefont {Shokeen}}, \bibinfo {author} {\bibfnamefont {M.}~\bibnamefont {Sanchez~Piaia}}, \bibinfo {author} {\bibfnamefont {J.-Y.}\ \bibnamefont {Bigot}}, \bibinfo {author} {\bibfnamefont {T.}~\bibnamefont {M\"uller}}, \bibinfo {author} {\bibfnamefont {P.}~\bibnamefont {Elliott}}, \bibinfo {author} {\bibfnamefont {J.~K.}\ \bibnamefont {Dewhurst}}, \bibinfo {author} {\bibfnamefont {S.}~\bibnamefont {Sharma}},\ and\ \bibinfo {author} {\bibfnamefont {E.~K.~U.}\ \bibnamefont {Gross}},\ }\bibfield  {title} {\bibinfo {title} {Spin flips versus spin transport in nonthermal electrons excited by ultrashort optical pulses in transition metals},\ }\href {https://doi.org/10.1103/PhysRevLett.119.107203} {\bibfield  {journal} {\bibinfo  {journal} {Phys. Rev. Lett.}\ }\textbf {\bibinfo {volume} {119}},\ \bibinfo {pages} {107203} (\bibinfo {year} {2017})}\BibitemShut {NoStop}%
	\bibitem [{\citenamefont {Chen}\ and\ \citenamefont {Wang}(2019)}]{Chen2019a}%
	\BibitemOpen
	\bibfield  {author} {\bibinfo {author} {\bibfnamefont {Z.}~\bibnamefont {Chen}}\ and\ \bibinfo {author} {\bibfnamefont {L.-W.}\ \bibnamefont {Wang}},\ }\bibfield  {title} {\bibinfo {title} {Role of initial magnetic disorder: A time-dependent {\it ab initio} study of ultrafast demagnetization mechanisms},\ }\href {https://doi.org/10.1126/sciadv.aau8000} {\bibfield  {journal} {\bibinfo  {journal} {Sci. Adv.}\ }\textbf {\bibinfo {volume} {5}},\ \bibinfo {pages} {eaau800} (\bibinfo {year} {2019})}\BibitemShut {NoStop}%
	\bibitem [{\citenamefont {Pellegrini}\ \emph {et~al.}(2022)\citenamefont {Pellegrini}, \citenamefont {Sharma}, \citenamefont {Dewhurst},\ and\ \citenamefont {Sanna}}]{Pellegrini2022}%
	\BibitemOpen
	\bibfield  {author} {\bibinfo {author} {\bibfnamefont {C.}~\bibnamefont {Pellegrini}}, \bibinfo {author} {\bibfnamefont {S.}~\bibnamefont {Sharma}}, \bibinfo {author} {\bibfnamefont {J.~K.}\ \bibnamefont {Dewhurst}},\ and\ \bibinfo {author} {\bibfnamefont {A.}~\bibnamefont {Sanna}},\ }\bibfield  {title} {\bibinfo {title} {{\it Ab initio} study of ultrafast demagnetization of elementary ferromagnets by terahertz versus optical pulses},\ }\href {https://doi.org/10.1103/PhysRevB.105.134425} {\bibfield  {journal} {\bibinfo  {journal} {Phys. Rev. B}\ }\textbf {\bibinfo {volume} {105}},\ \bibinfo {pages} {134425} (\bibinfo {year} {2022})}\BibitemShut {NoStop}%
	\bibitem [{\citenamefont {Mrudul}\ and\ \citenamefont {Oppeneer}(2024)}]{Mrudul2024}%
	\BibitemOpen
	\bibfield  {author} {\bibinfo {author} {\bibfnamefont {M.~S.}\ \bibnamefont {Mrudul}}\ and\ \bibinfo {author} {\bibfnamefont {P.~M.}\ \bibnamefont {Oppeneer}},\ }\bibfield  {title} {\bibinfo {title} {{\em Ab initio} investigation of laser-induced ultrafast demagnetization of {$L{1}_{0}$ FePt}: Intensity dependence and importance of electron coherence},\ }\href {https://doi.org/10.1103/PhysRevB.109.144418} {\bibfield  {journal} {\bibinfo  {journal} {Phys. Rev. B}\ }\textbf {\bibinfo {volume} {109}},\ \bibinfo {pages} {144418} (\bibinfo {year} {2024})}\BibitemShut {NoStop}%
	\bibitem [{\citenamefont {Kefayati}\ and\ \citenamefont {Nikoli\'{c}}(2024)}]{Kefayati2024}%
	\BibitemOpen
	\bibfield  {author} {\bibinfo {author} {\bibfnamefont {A.}~\bibnamefont {Kefayati}}\ and\ \bibinfo {author} {\bibfnamefont {B.~K.}\ \bibnamefont {Nikoli\'{c}}},\ }\bibfield  {title} {\bibinfo {title} {Origins of electromagnetic radiation from spintronic terahertz emitters: A time-dependent density functional theory plus {Jefimenko} equations approach},\ }\href {https://doi.org/10.1103/PhysRevLett.133.136704} {\bibfield  {journal} {\bibinfo  {journal} {Phys. Rev. Lett.}\ }\textbf {\bibinfo {volume} {133}},\ \bibinfo {pages} {136704} (\bibinfo {year} {2024})}\BibitemShut {NoStop}%
	\bibitem [{\citenamefont {Tancogne-Dejean}\ \emph {et~al.}(2022)\citenamefont {Tancogne-Dejean}, \citenamefont {Eich},\ and\ \citenamefont {Rubio}}]{Dejean2022}%
	\BibitemOpen
	\bibfield  {author} {\bibinfo {author} {\bibfnamefont {N.}~\bibnamefont {Tancogne-Dejean}}, \bibinfo {author} {\bibfnamefont {F.~G.}\ \bibnamefont {Eich}},\ and\ \bibinfo {author} {\bibfnamefont {A.}~\bibnamefont {Rubio}},\ }\bibfield  {title} {\bibinfo {title} {Effect of spin-orbit coupling on the high harmonics from the topological dirac semimetal {Na$_3$Bi}},\ }\href {https://doi.org/10.1038/s41524-022-00831-6} {\bibfield  {journal} {\bibinfo  {journal} {Npj Comput. Mater.}\ }\textbf {\bibinfo {volume} {8}},\ \bibinfo {pages} {145} (\bibinfo {year} {2022})}\BibitemShut {NoStop}%
	\bibitem [{\citenamefont {Tserkovnyak}\ \emph {et~al.}(2005)\citenamefont {Tserkovnyak}, \citenamefont {Brataas}, \citenamefont {Bauer},\ and\ \citenamefont {Halperin}}]{Tserkovnyak2005}%
	\BibitemOpen
	\bibfield  {author} {\bibinfo {author} {\bibfnamefont {Y.}~\bibnamefont {Tserkovnyak}}, \bibinfo {author} {\bibfnamefont {A.}~\bibnamefont {Brataas}}, \bibinfo {author} {\bibfnamefont {G.~E.}\ \bibnamefont {Bauer}},\ and\ \bibinfo {author} {\bibfnamefont {B.~I.}\ \bibnamefont {Halperin}},\ }\bibfield  {title} {\bibinfo {title} {Nonlocal magnetization dynamics in ferromagnetic heterostructures},\ }\href {https://doi.org/10.1103/RevModPhys.77.1375} {\bibfield  {journal} {\bibinfo  {journal} {Rev. Mod. Phys.}\ }\textbf {\bibinfo {volume} {77}},\ \bibinfo {pages} {1375} (\bibinfo {year} {2005})}\BibitemShut {NoStop}%
	\bibitem [{\citenamefont {Ando}(2014)}]{Ando2014a}%
	\BibitemOpen
	\bibfield  {author} {\bibinfo {author} {\bibfnamefont {K.}~\bibnamefont {Ando}},\ }\bibfield  {title} {\bibinfo {title} {Dynamical generation of spin currents},\ }\href {https://doi.org/10.1088/0268-1242/29/4/043002} {\bibfield  {journal} {\bibinfo  {journal} {Semicond. Sci. Technol.}\ }\textbf {\bibinfo {volume} {29}},\ \bibinfo {pages} {043002} (\bibinfo {year} {2014})}\BibitemShut {NoStop}%
	\bibitem [{\citenamefont {Garcia-Gaitan}\ \emph {et~al.}(2025)\citenamefont {Garcia-Gaitan}, \citenamefont {Feiguin},\ and\ \citenamefont {Nikolic}}]{GarciaGaitan2025a}%
	\BibitemOpen
	\bibfield  {author} {\bibinfo {author} {\bibfnamefont {F.}~\bibnamefont {Garcia-Gaitan}}, \bibinfo {author} {\bibfnamefont {A.~E.}\ \bibnamefont {Feiguin}},\ and\ \bibinfo {author} {\bibfnamefont {B.~K.}\ \bibnamefont {Nikolic}},\ }\bibfield  {title} {\bibinfo {title} {Nonclassical dynamics of {N\'eel} vector and magnetization accompanied by {THz} and high-harmonic radiation from ultrafast-light-driven antiferromagnetic {Mott} insulator},\ }\href {https://arxiv.org/abs/2502.00849} {\bibfield  {journal} {\bibinfo  {journal} {arXiv:2502.00849}\ } (\bibinfo {year} {2025})}\BibitemShut {NoStop}%
	\bibitem [{\citenamefont {Citro}\ and\ \citenamefont {Aidelsburger}(2023)}]{Citro2023}%
	\BibitemOpen
	\bibfield  {author} {\bibinfo {author} {\bibfnamefont {R.}~\bibnamefont {Citro}}\ and\ \bibinfo {author} {\bibfnamefont {M.}~\bibnamefont {Aidelsburger}},\ }\bibfield  {title} {\bibinfo {title} {Thouless pumping and topology},\ }\href {https://doi.org/10.1038/s42254-022-00545-0} {\bibfield  {journal} {\bibinfo  {journal} {Nat. Rev. Phys.}\ }\textbf {\bibinfo {volume} {5}},\ \bibinfo {pages} {87} (\bibinfo {year} {2023})}\BibitemShut {NoStop}%
	\bibitem [{\citenamefont {Brouwer}(1998)}]{Brouwer1998}%
	\BibitemOpen
	\bibfield  {author} {\bibinfo {author} {\bibfnamefont {P.~W.}\ \bibnamefont {Brouwer}},\ }\bibfield  {title} {\bibinfo {title} {Scattering approach to parametric pumping},\ }\href {https://doi.org/10.1103/PhysRevB.58.R10135} {\bibfield  {journal} {\bibinfo  {journal} {Phys. Rev. B}\ }\textbf {\bibinfo {volume} {58}},\ \bibinfo {pages} {R10135} (\bibinfo {year} {1998})}\BibitemShut {NoStop}%
	\bibitem [{\citenamefont {Varela-Manjarres}\ and\ \citenamefont {Nikoli\'{c}}(2023)}]{VarelaManjarres2023}%
	\BibitemOpen
	\bibfield  {author} {\bibinfo {author} {\bibfnamefont {J.}~\bibnamefont {Varela-Manjarres}}\ and\ \bibinfo {author} {\bibfnamefont {B.~K.}\ \bibnamefont {Nikoli\'{c}}},\ }\bibfield  {title} {\bibinfo {title} {High-harmonic generation in spin and charge current pumping at ferromagnetic or antiferromagnetic resonance in the presence of spin-orbit coupling},\ }\href {https://doi.org/10.1088/2515-7639/aceaad} {\bibfield  {journal} {\bibinfo  {journal} {J. Phys.: Mater.}\ }\textbf {\bibinfo {volume} {6}},\ \bibinfo {pages} {045001} (\bibinfo {year} {2023})}\BibitemShut {NoStop}%
	\bibitem [{Note3()}]{Note3}%
	\BibitemOpen
	\bibinfo {note} {It is worth recalling arguments for how terminology ``pumping'' has been attached~\cite {Freimuth2017} to wider and wider classes of phenomena, while keeping them unified through common equations describing apparently vastly different situations, which we also follow here. Adiabatic quantum pumping was introduced by Thouless~\cite {Citro2023} as an effect, in which slow modulation of two or more external parameters of a quantum system, results in a net DC charge current in the absence of any externally applied bias voltage. In particular, Brouwer scattering-matrix-based formula~\cite {Brouwer1998} for such charge pumping from a quantum dot, attached to two leads and modulated by two periodically changing gate voltages as fabricated experimentally~\cite {Switkes1999}, was later applied~\cite {Tserkovnyak2005} to FM/NM bilayers to explain why precessing magnetization of FM layer generates pure (i.e., with no accompanied charge current) spin current flowing toward NM layer. In this approach,
		two time-periodic components of magnetization (due to precession; the third component is fixed along the axis of precession) replace two gate voltages of the original Brouwer scattering-matrix-based formula~\cite {Brouwer1998}. Since the same formula explains both phenomena, such spin current generation is legitimately also termed~\cite {Tserkovnyak2005} ``spin pumping.'' The same effect can be described by more general Keldysh GFs-based expressions~\cite {Mahfouzi2012,VarelaManjarres2023,Dolui2020b}, which can also handle situations when magnetization changes non-periodically~\cite {Petrovic2018,Petrovic2021,Abbout2018} or is changing length~\cite {VarelaManjarres2024} instead of precessing. In the latter case, identical system---such as an illustrative example~\cite {Chen2009} of a single site with magnetization vector within one-dimensional tigh-binding chain---treated by the same time-dependent Keldysh GFs-based calculations will produce spin current~\cite {Chen2009,Petrovic2018} if magnetization vector
		is precessing (which is identical~\cite {Chen2009} to the one~\cite {Tserkovnyak2005} computed from the Brouwer scattering-matrix-based formula~\cite {Brouwer1998}), or it will produce both~\cite {VarelaManjarres2024} spin and charge currents if its length is shrinking aperiodically~\cite {Freimuth2017} (as in demagnetization). This justifies usage of ``pumping''~\cite {Freimuth2017} of spin and charge terminology for all such effects.}\BibitemShut {Stop}%
	\bibitem [{\citenamefont {Choi}\ \emph {et~al.}(2014)\citenamefont {Choi}, \citenamefont {Min}, \citenamefont {Lee},\ and\ \citenamefont {Cahill}}]{Choi2014}%
	\BibitemOpen
	\bibfield  {author} {\bibinfo {author} {\bibfnamefont {G.-M.}\ \bibnamefont {Choi}}, \bibinfo {author} {\bibfnamefont {B.-C.}\ \bibnamefont {Min}}, \bibinfo {author} {\bibfnamefont {K.-J.}\ \bibnamefont {Lee}},\ and\ \bibinfo {author} {\bibfnamefont {D.~G.}\ \bibnamefont {Cahill}},\ }\bibfield  {title} {\bibinfo {title} {Spin current generated by thermally driven ultrafast demagnetization},\ }\href {https://doi.org/10.1038/ncomms5334} {\bibfield  {journal} {\bibinfo  {journal} {Nat. Commun.}\ }\textbf {\bibinfo {volume} {5}},\ \bibinfo {pages} {4334} (\bibinfo {year} {2014})}\BibitemShut {NoStop}%
	\bibitem [{\citenamefont {Lichtenberg}\ \emph {et~al.}(2022)\citenamefont {Lichtenberg}, \citenamefont {Beens}, \citenamefont {Jansen}, \citenamefont {Koopmans},\ and\ \citenamefont {Duine}}]{Lichtenberg2022}%
	\BibitemOpen
	\bibfield  {author} {\bibinfo {author} {\bibfnamefont {T.}~\bibnamefont {Lichtenberg}}, \bibinfo {author} {\bibfnamefont {M.}~\bibnamefont {Beens}}, \bibinfo {author} {\bibfnamefont {M.~H.}\ \bibnamefont {Jansen}}, \bibinfo {author} {\bibfnamefont {B.}~\bibnamefont {Koopmans}},\ and\ \bibinfo {author} {\bibfnamefont {R.~A.}\ \bibnamefont {Duine}},\ }\bibfield  {title} {\bibinfo {title} {Probing optically induced spin currents using terahertz spin waves in noncollinear magnetic bilayers},\ }\href {https://doi.org/10.1103/PhysRevB.105.144416} {\bibfield  {journal} {\bibinfo  {journal} {Phys. Rev. B}\ }\textbf {\bibinfo {volume} {105}},\ \bibinfo {pages} {144416} (\bibinfo {year} {2022})}\BibitemShut {NoStop}%
	\bibitem [{\citenamefont {Marmolejo-Tejada}\ \emph {et~al.}(2017)\citenamefont {Marmolejo-Tejada}, \citenamefont {Dolui}, \citenamefont {Lazi\'{c}}, \citenamefont {Chang}, \citenamefont {Smidstrup}, \citenamefont {Stradi}, \citenamefont {Stokbro},\ and\ \citenamefont {Nikoli\'{c}}}]{MarmolejoTejada2017}%
	\BibitemOpen
	\bibfield  {author} {\bibinfo {author} {\bibfnamefont {J.~M.}\ \bibnamefont {Marmolejo-Tejada}}, \bibinfo {author} {\bibfnamefont {K.}~\bibnamefont {Dolui}}, \bibinfo {author} {\bibfnamefont {P.}~\bibnamefont {Lazi\'{c}}}, \bibinfo {author} {\bibfnamefont {P.-H.}\ \bibnamefont {Chang}}, \bibinfo {author} {\bibfnamefont {S.}~\bibnamefont {Smidstrup}}, \bibinfo {author} {\bibfnamefont {D.}~\bibnamefont {Stradi}}, \bibinfo {author} {\bibfnamefont {K.}~\bibnamefont {Stokbro}},\ and\ \bibinfo {author} {\bibfnamefont {B.~K.}\ \bibnamefont {Nikoli\'{c}}},\ }\bibfield  {title} {\bibinfo {title} {Proximity band structure and spin textures on both sides of topological-insulator/ferromagnetic-metal interface and their charge transport probes},\ }\href {https://doi.org/10.1021/acs.nanolett.7b02511} {\bibfield  {journal} {\bibinfo  {journal} {Nano Lett.}\ }\textbf {\bibinfo {volume} {17}},\ \bibinfo {pages} {5626} (\bibinfo {year} {2017})}\BibitemShut {NoStop}%
	\bibitem [{\citenamefont {Dolui}\ and\ \citenamefont {Nikoli{\'{c}}}(2020)}]{Dolui2020a}%
	\BibitemOpen
	\bibfield  {author} {\bibinfo {author} {\bibfnamefont {K.}~\bibnamefont {Dolui}}\ and\ \bibinfo {author} {\bibfnamefont {B.~K.}\ \bibnamefont {Nikoli{\'{c}}}},\ }\bibfield  {title} {\bibinfo {title} {Spin-orbit-proximitized ferromagnetic metal by monolayer transition metal dichalcogenide: {Atlas} of spectral functions, spin textures, and spin-orbit torques in {Co/MoSe$_2$}, {Co/WSe$_2$}, and {Co/TaSe$_2$} heterostructures},\ }\href {https://doi.org/10.1103/physrevmaterials.4.104007} {\bibfield  {journal} {\bibinfo  {journal} {Phys. Rev. Mater.}\ }\textbf {\bibinfo {volume} {4}},\ \bibinfo {pages} {104007} (\bibinfo {year} {2020})}\BibitemShut {NoStop}%
	\bibitem [{\citenamefont {\v{Z}uti\'{c}}\ \emph {et~al.}(2019)\citenamefont {\v{Z}uti\'{c}}, \citenamefont {Matos-Abiague}, \citenamefont {Scharf}, \citenamefont {Dery},\ and\ \citenamefont {Belashchenko}}]{Zutic2019}%
	\BibitemOpen
	\bibfield  {author} {\bibinfo {author} {\bibfnamefont {I.}~\bibnamefont {\v{Z}uti\'{c}}}, \bibinfo {author} {\bibfnamefont {A.}~\bibnamefont {Matos-Abiague}}, \bibinfo {author} {\bibfnamefont {B.}~\bibnamefont {Scharf}}, \bibinfo {author} {\bibfnamefont {H.}~\bibnamefont {Dery}},\ and\ \bibinfo {author} {\bibfnamefont {K.}~\bibnamefont {Belashchenko}},\ }\bibfield  {title} {\bibinfo {title} {Proximitized materials},\ }\href {https://doi.org/https://doi.org/10.1016/j.mattod.2018.05.003} {\bibfield  {journal} {\bibinfo  {journal} {Mater. Today}\ }\textbf {\bibinfo {volume} {22}},\ \bibinfo {pages} {85 } (\bibinfo {year} {2019})}\BibitemShut {NoStop}%
	\bibitem [{\citenamefont {Kuiper}\ \emph {et~al.}(2014)\citenamefont {Kuiper}, \citenamefont {Roth}, \citenamefont {Schellekens}, \citenamefont {Schmitt}, \citenamefont {Koopmans}, \citenamefont {Cinchetti},\ and\ \citenamefont {Aeschlimann}}]{Kuiper2014}%
	\BibitemOpen
	\bibfield  {author} {\bibinfo {author} {\bibfnamefont {K.}~\bibnamefont {Kuiper}}, \bibinfo {author} {\bibfnamefont {T.}~\bibnamefont {Roth}}, \bibinfo {author} {\bibfnamefont {A.}~\bibnamefont {Schellekens}}, \bibinfo {author} {\bibfnamefont {O.}~\bibnamefont {Schmitt}}, \bibinfo {author} {\bibfnamefont {B.}~\bibnamefont {Koopmans}}, \bibinfo {author} {\bibfnamefont {M.}~\bibnamefont {Cinchetti}},\ and\ \bibinfo {author} {\bibfnamefont {M.}~\bibnamefont {Aeschlimann}},\ }\bibfield  {title} {\bibinfo {title} {Spin-orbit enhanced demagnetization rate in {C}o/{P}t-multilayers},\ }\href {https://doi.org/https://doi.org/10.1063/1.4902069} {\bibfield  {journal} {\bibinfo  {journal} {Appl. Phys. Lett.}\ }\textbf {\bibinfo {volume} {105}},\ \bibinfo {pages} {1} (\bibinfo {year} {2014})}\BibitemShut {NoStop}%
	\bibitem [{\citenamefont {Malinowski}\ \emph {et~al.}(2008)\citenamefont {Malinowski}, \citenamefont {Longa}, \citenamefont {Rietjens}, \citenamefont {Paluskar}, \citenamefont {Huijink}, \citenamefont {Swagten},\ and\ \citenamefont {Koopmans}}]{Malinowski2008}%
	\BibitemOpen
	\bibfield  {author} {\bibinfo {author} {\bibfnamefont {G.}~\bibnamefont {Malinowski}}, \bibinfo {author} {\bibfnamefont {F.~D.}\ \bibnamefont {Longa}}, \bibinfo {author} {\bibfnamefont {J.~H.~H.}\ \bibnamefont {Rietjens}}, \bibinfo {author} {\bibfnamefont {P.~V.}\ \bibnamefont {Paluskar}}, \bibinfo {author} {\bibfnamefont {R.}~\bibnamefont {Huijink}}, \bibinfo {author} {\bibfnamefont {H.~J.~M.}\ \bibnamefont {Swagten}},\ and\ \bibinfo {author} {\bibfnamefont {B.}~\bibnamefont {Koopmans}},\ }\bibfield  {title} {\bibinfo {title} {Control of speed and efficiency of ultrafast demagnetization by direct transfer of spin angular momentum},\ }\href {https://doi.org/10.1038/nphys1092} {\bibfield  {journal} {\bibinfo  {journal} {Nat. Phys.}\ }\textbf {\bibinfo {volume} {4}},\ \bibinfo {pages} {855} (\bibinfo {year} {2008})}\BibitemShut {NoStop}%
	\bibitem [{\citenamefont {Suresh}\ and\ \citenamefont {Nikoli\ifmmode~\acute{c}\else \'{c}\fi{}}(2023)}]{Suresh2023}%
	\BibitemOpen
	\bibfield  {author} {\bibinfo {author} {\bibfnamefont {A.}~\bibnamefont {Suresh}}\ and\ \bibinfo {author} {\bibfnamefont {B.~K.}\ \bibnamefont {Nikoli\ifmmode~\acute{c}\else \'{c}\fi{}}},\ }\bibfield  {title} {\bibinfo {title} {Quantum classical approach to spin and charge pumping and the ensuing radiation in terahertz spintronics: Example of the ultrafast light-driven {Weyl} antiferromagnet $\text{Mn}_{3}\text{Sn}$},\ }\href {https://doi.org/10.1103/PhysRevB.107.174421} {\bibfield  {journal} {\bibinfo  {journal} {Phys. Rev. B}\ }\textbf {\bibinfo {volume} {107}},\ \bibinfo {pages} {174421} (\bibinfo {year} {2023})}\BibitemShut {NoStop}%
	\bibitem [{\citenamefont {Bajpai}\ \emph {et~al.}(2019)\citenamefont {Bajpai}, \citenamefont {Popescu}, \citenamefont {Plech\'a\v{c}}, \citenamefont {Nikoli\'{c}}, \citenamefont {Torres}, \citenamefont {Ishizuka},\ and\ \citenamefont {Nagaosa}}]{Bajpai2019}%
	\BibitemOpen
	\bibfield  {author} {\bibinfo {author} {\bibfnamefont {U.}~\bibnamefont {Bajpai}}, \bibinfo {author} {\bibfnamefont {B.~S.}\ \bibnamefont {Popescu}}, \bibinfo {author} {\bibfnamefont {P.}~\bibnamefont {Plech\'a\v{c}}}, \bibinfo {author} {\bibfnamefont {B.~K.}\ \bibnamefont {Nikoli\'{c}}}, \bibinfo {author} {\bibfnamefont {L.~E. F.~F.}\ \bibnamefont {Torres}}, \bibinfo {author} {\bibfnamefont {H.}~\bibnamefont {Ishizuka}},\ and\ \bibinfo {author} {\bibfnamefont {N.}~\bibnamefont {Nagaosa}},\ }\bibfield  {title} {\bibinfo {title} {Spatio-temporal dynamics of shift current quantum pumping by femtosecond light pulse},\ }\href {https://doi.org/10.1088/2515-7639/ab0a3e} {\bibfield  {journal} {\bibinfo  {journal} {J. Phys.: Mater.}\ }\textbf {\bibinfo {volume} {2}},\ \bibinfo {pages} {025004} (\bibinfo {year} {2019})}\BibitemShut {NoStop}%
	\bibitem [{\citenamefont {Agarwal}\ \emph {et~al.}(2023)\citenamefont {Agarwal}, \citenamefont {Yang}, \citenamefont {Medwal}, \citenamefont {Asada}, \citenamefont {Fukuma}, \citenamefont {Battiato},\ and\ \citenamefont {Singh}}]{Agarwal2023}%
	\BibitemOpen
	\bibfield  {author} {\bibinfo {author} {\bibfnamefont {P.}~\bibnamefont {Agarwal}}, \bibinfo {author} {\bibfnamefont {Y.}~\bibnamefont {Yang}}, \bibinfo {author} {\bibfnamefont {R.}~\bibnamefont {Medwal}}, \bibinfo {author} {\bibfnamefont {H.}~\bibnamefont {Asada}}, \bibinfo {author} {\bibfnamefont {Y.}~\bibnamefont {Fukuma}}, \bibinfo {author} {\bibfnamefont {M.}~\bibnamefont {Battiato}},\ and\ \bibinfo {author} {\bibfnamefont {R.}~\bibnamefont {Singh}},\ }\bibfield  {title} {\bibinfo {title} {Secondary spin current driven efficient {THz} spintronic emitters},\ }\href {https://doi.org/10.1002/adom.202301027} {\bibfield  {journal} {\bibinfo  {journal} {Adv. Opt. Mater.}\ }\textbf {\bibinfo {volume} {11}},\ \bibinfo {pages} {2301027} (\bibinfo {year} {2023})}\BibitemShut {NoStop}%
	\bibitem [{\citenamefont {Beaurepaire}\ \emph {et~al.}(2004)\citenamefont {Beaurepaire}, \citenamefont {Turner}, \citenamefont {Harrel}, \citenamefont {Beard}, \citenamefont {Bigot},\ and\ \citenamefont {Schmuttenmaer}}]{Beaurepaire2004}%
	\BibitemOpen
	\bibfield  {author} {\bibinfo {author} {\bibfnamefont {E.}~\bibnamefont {Beaurepaire}}, \bibinfo {author} {\bibfnamefont {G.~M.}\ \bibnamefont {Turner}}, \bibinfo {author} {\bibfnamefont {S.~M.}\ \bibnamefont {Harrel}}, \bibinfo {author} {\bibfnamefont {M.~C.}\ \bibnamefont {Beard}}, \bibinfo {author} {\bibfnamefont {J.-Y.}\ \bibnamefont {Bigot}},\ and\ \bibinfo {author} {\bibfnamefont {C.~A.}\ \bibnamefont {Schmuttenmaer}},\ }\bibfield  {title} {\bibinfo {title} {Coherent terahertz emission from ferromagnetic films excited by femtosecond laser pulses},\ }\href {https://doi.org/10.1063/1.1737467} {\bibfield  {journal} {\bibinfo  {journal} {Appl. Phys. Lett.}\ }\textbf {\bibinfo {volume} {84}},\ \bibinfo {pages} {3465} (\bibinfo {year} {2004})}\BibitemShut {NoStop}%
	\bibitem [{\citenamefont {Jefimenko}(1966)}]{Jefimenko1966}%
	\BibitemOpen
	\bibfield  {author} {\bibinfo {author} {\bibfnamefont {O.~D.}\ \bibnamefont {Jefimenko}},\ }\href@noop {} {\emph {\bibinfo {title} {Electricity and Magnetism}}}\ (\bibinfo  {publisher} {Appleton Century-Crofts, New York},\ \bibinfo {year} {1966})\BibitemShut {NoStop}%
	\bibitem [{\citenamefont {McDonald}(1997)}]{McDonald1997}%
	\BibitemOpen
	\bibfield  {author} {\bibinfo {author} {\bibfnamefont {K.~T.}\ \bibnamefont {McDonald}},\ }\bibfield  {title} {\bibinfo {title} {The relation between expressions for time-dependent electromagnetic fields given by {Jefimenko} and by {Panofsky} and {Phillips}},\ }\href {https://doi.org/10.1119/1.18723} {\bibfield  {journal} {\bibinfo  {journal} {Am. J. Phys.}\ }\textbf {\bibinfo {volume} {65}},\ \bibinfo {pages} {1074} (\bibinfo {year} {1997})}\BibitemShut {NoStop}%
	\bibitem [{\citenamefont {Griffiths}\ and\ \citenamefont {Heald}(1991)}]{Griffiths1991}%
	\BibitemOpen
	\bibfield  {author} {\bibinfo {author} {\bibfnamefont {D.~J.}\ \bibnamefont {Griffiths}}\ and\ \bibinfo {author} {\bibfnamefont {M.~A.}\ \bibnamefont {Heald}},\ }\bibfield  {title} {\bibinfo {title} {Time-dependent generalizations of the {Biot–Savart} and {Coulomb} laws},\ }\href {https://doi.org/10.1119/1.16589} {\bibfield  {journal} {\bibinfo  {journal} {Am. J. Phys.}\ }\textbf {\bibinfo {volume} {59}},\ \bibinfo {pages} {111} (\bibinfo {year} {1991})}\BibitemShut {NoStop}%
	\bibitem [{\citenamefont {Tancogne-Dejean}\ \emph {et~al.}(2020)\citenamefont {Tancogne-Dejean}, \citenamefont {Oliveira}, \citenamefont {Andrade}, \citenamefont {Appel}, \citenamefont {Borca}, \citenamefont {Breton}, \citenamefont {Buchholz}, \citenamefont {Castro}, \citenamefont {Corni}, \citenamefont {Correa} \emph {et~al.}}]{Tancogne-Dejean2020}%
	\BibitemOpen
	\bibfield  {author} {\bibinfo {author} {\bibfnamefont {N.}~\bibnamefont {Tancogne-Dejean}}, \bibinfo {author} {\bibfnamefont {M.~J.~T.}\ \bibnamefont {Oliveira}}, \bibinfo {author} {\bibfnamefont {X.}~\bibnamefont {Andrade}}, \bibinfo {author} {\bibfnamefont {H.}~\bibnamefont {Appel}}, \bibinfo {author} {\bibfnamefont {C.~H.}\ \bibnamefont {Borca}}, \bibinfo {author} {\bibfnamefont {G.~L.}\ \bibnamefont {Breton}}, \bibinfo {author} {\bibfnamefont {F.}~\bibnamefont {Buchholz}}, \bibinfo {author} {\bibfnamefont {A.}~\bibnamefont {Castro}}, \bibinfo {author} {\bibfnamefont {S.}~\bibnamefont {Corni}}, \bibinfo {author} {\bibfnamefont {A.~A.}\ \bibnamefont {Correa}}, \emph {et~al.},\ }\bibfield  {title} {\bibinfo {title} {Octopus, a computational framework for exploring light-driven phenomena and quantum dynamics in extended and finite systems},\ }\href {https://doi.org/10.1063/1.5142502} {\bibfield  {journal} {\bibinfo  {journal} {J. Chem. Phys.}\ }\textbf {\bibinfo {volume} {152}},\ \bibinfo {pages} {124119}
		(\bibinfo {year} {2020})}\BibitemShut {NoStop}%
	\bibitem [{\citenamefont {Philip}\ and\ \citenamefont {Gilbert}(2018)}]{Philip2018}%
	\BibitemOpen
	\bibfield  {author} {\bibinfo {author} {\bibfnamefont {T.~M.}\ \bibnamefont {Philip}}\ and\ \bibinfo {author} {\bibfnamefont {M.~J.}\ \bibnamefont {Gilbert}},\ }\bibfield  {title} {\bibinfo {title} {Theory of {AC} quantum transport with fully electrodynamic coupling},\ }\href {https://doi.org/10.1007/s10825-018-1191-z} {\bibfield  {journal} {\bibinfo  {journal} {J. Comput. Electron.}\ }\textbf {\bibinfo {volume} {17}},\ \bibinfo {pages} {934} (\bibinfo {year} {2018})}\BibitemShut {NoStop}%
	\bibitem [{\citenamefont {Mahfouzi}\ \emph {et~al.}(2012)\citenamefont {Mahfouzi}, \citenamefont {Fabian}, \citenamefont {Nagaosa},\ and\ \citenamefont {Nikoli\ifmmode~\acute{c}\else \'{c}\fi{}}}]{Mahfouzi2012}%
	\BibitemOpen
	\bibfield  {author} {\bibinfo {author} {\bibfnamefont {F.}~\bibnamefont {Mahfouzi}}, \bibinfo {author} {\bibfnamefont {J.}~\bibnamefont {Fabian}}, \bibinfo {author} {\bibfnamefont {N.}~\bibnamefont {Nagaosa}},\ and\ \bibinfo {author} {\bibfnamefont {B.~K.}\ \bibnamefont {Nikoli\ifmmode~\acute{c}\else \'{c}\fi{}}},\ }\bibfield  {title} {\bibinfo {title} {Charge pumping by magnetization dynamics in magnetic and semimagnetic tunnel junctions with interfacial {Rashba} or bulk extrinsic spin-orbit coupling},\ }\href {https://doi.org/10.1103/PhysRevB.85.054406} {\bibfield  {journal} {\bibinfo  {journal} {Phys. Rev. B}\ }\textbf {\bibinfo {volume} {85}},\ \bibinfo {pages} {054406} (\bibinfo {year} {2012})}\BibitemShut {NoStop}%
	\bibitem [{\citenamefont {Volkov}\ \emph {et~al.}(2019)\citenamefont {Volkov}, \citenamefont {Sato}, \citenamefont {Schlaepfer}, \citenamefont {Kasmi}, \citenamefont {Hartmann}, \citenamefont {Lucchini}, \citenamefont {Gallmann}, \citenamefont {Rubio},\ and\ \citenamefont {Keller}}]{Volkov2019}%
	\BibitemOpen
	\bibfield  {author} {\bibinfo {author} {\bibfnamefont {M.}~\bibnamefont {Volkov}}, \bibinfo {author} {\bibfnamefont {S.~A.}\ \bibnamefont {Sato}}, \bibinfo {author} {\bibfnamefont {F.}~\bibnamefont {Schlaepfer}}, \bibinfo {author} {\bibfnamefont {L.}~\bibnamefont {Kasmi}}, \bibinfo {author} {\bibfnamefont {N.}~\bibnamefont {Hartmann}}, \bibinfo {author} {\bibfnamefont {M.}~\bibnamefont {Lucchini}}, \bibinfo {author} {\bibfnamefont {L.}~\bibnamefont {Gallmann}}, \bibinfo {author} {\bibfnamefont {A.}~\bibnamefont {Rubio}},\ and\ \bibinfo {author} {\bibfnamefont {U.}~\bibnamefont {Keller}},\ }\bibfield  {title} {\bibinfo {title} {Attosecond screening dynamics mediated by electron localization in transition metals},\ }\href {https://doi.org/10.1038/s41567-019-0602-9} {\bibfield  {journal} {\bibinfo  {journal} {Nat. Phys.}\ }\textbf {\bibinfo {volume} {15}},\ \bibinfo {pages} {1145} (\bibinfo {year} {2019})}\BibitemShut {NoStop}%
	\bibitem [{\citenamefont {de~Vos}\ \emph {et~al.}(2023)\citenamefont {de~Vos}, \citenamefont {Neb}, \citenamefont {Niedermayr}, \citenamefont {Burri}, \citenamefont {Hollm}, \citenamefont {Gallmann},\ and\ \citenamefont {Keller}}]{Vos2023}%
	\BibitemOpen
	\bibfield  {author} {\bibinfo {author} {\bibfnamefont {E.~W.}\ \bibnamefont {de~Vos}}, \bibinfo {author} {\bibfnamefont {S.}~\bibnamefont {Neb}}, \bibinfo {author} {\bibfnamefont {A.}~\bibnamefont {Niedermayr}}, \bibinfo {author} {\bibfnamefont {F.}~\bibnamefont {Burri}}, \bibinfo {author} {\bibfnamefont {M.}~\bibnamefont {Hollm}}, \bibinfo {author} {\bibfnamefont {L.}~\bibnamefont {Gallmann}},\ and\ \bibinfo {author} {\bibfnamefont {U.}~\bibnamefont {Keller}},\ }\bibfield  {title} {\bibinfo {title} {Ultrafast transition from state-blocking dynamics to electron localization in transition metal $\ensuremath{\beta}$-tungsten},\ }\href {https://doi.org/10.1103/PhysRevLett.131.226901} {\bibfield  {journal} {\bibinfo  {journal} {Phys. Rev. Lett.}\ }\textbf {\bibinfo {volume} {131}},\ \bibinfo {pages} {226901} (\bibinfo {year} {2023})}\BibitemShut {NoStop}%
	\bibitem [{\citenamefont {Adamantopoulos}\ \emph {et~al.}(2022)\citenamefont {Adamantopoulos}, \citenamefont {Merte}, \citenamefont {Go}, \citenamefont {Freimuth}, \citenamefont {Bl\"ugel},\ and\ \citenamefont {Mokrousov}}]{Adamantopoulos2022}%
	\BibitemOpen
	\bibfield  {author} {\bibinfo {author} {\bibfnamefont {T.}~\bibnamefont {Adamantopoulos}}, \bibinfo {author} {\bibfnamefont {M.}~\bibnamefont {Merte}}, \bibinfo {author} {\bibfnamefont {D.}~\bibnamefont {Go}}, \bibinfo {author} {\bibfnamefont {F.}~\bibnamefont {Freimuth}}, \bibinfo {author} {\bibfnamefont {S.}~\bibnamefont {Bl\"ugel}},\ and\ \bibinfo {author} {\bibfnamefont {Y.}~\bibnamefont {Mokrousov}},\ }\bibfield  {title} {\bibinfo {title} {Laser-induced charge and spin photocurrents at the {${\mathrm{BiAg}}_{2}$} surface: {A} first-principles benchmark},\ }\href {https://doi.org/10.1103/PhysRevResearch.4.043046} {\bibfield  {journal} {\bibinfo  {journal} {Phys. Rev. Res.}\ }\textbf {\bibinfo {volume} {4}},\ \bibinfo {pages} {043046} (\bibinfo {year} {2022})}\BibitemShut {NoStop}%
	\bibitem [{elk()}]{elkjefimenko}%
	\BibitemOpen
	\href@noop {} {\bibinfo {title} {\url{https://wiki.physics.udel.edu/qttg/Download_Research_Software_by_QTTG}}}\BibitemShut {NoStop}%
	\bibitem [{\citenamefont {Dewhurst}\ \emph {et~al.}(2016)\citenamefont {Dewhurst}, \citenamefont {Krieger}, \citenamefont {Sharma},\ and\ \citenamefont {Gross}}]{Dewhurst2016}%
	\BibitemOpen
	\bibfield  {author} {\bibinfo {author} {\bibfnamefont {J.~K.}\ \bibnamefont {Dewhurst}}, \bibinfo {author} {\bibfnamefont {K.}~\bibnamefont {Krieger}}, \bibinfo {author} {\bibfnamefont {S.}~\bibnamefont {Sharma}},\ and\ \bibinfo {author} {\bibfnamefont {E.~K.~U.}\ \bibnamefont {Gross}},\ }\bibfield  {title} {\bibinfo {title} {An efficient algorithm for time propagation as applied to linearized augmented plane wave method},\ }\href {https://doi.org/10.1016/j.cpc.2016.09.001} {\bibfield  {journal} {\bibinfo  {journal} {Comput. Phys. Commun.}\ }\textbf {\bibinfo {volume} {209}},\ \bibinfo {pages} {92} (\bibinfo {year} {2016})}\BibitemShut {NoStop}%
	\bibitem [{\citenamefont {\url{http://elk.sourceforge.net/}}()}]{elk}%
	\BibitemOpen
	\bibfield  {author} {\bibinfo {author} {\bibnamefont {\url{http://elk.sourceforge.net/}}},\ }\href@noop {} {}\BibitemShut {NoStop}%
	\bibitem [{\citenamefont {Ullrich}(2011)}]{Ullrich2011}%
	\BibitemOpen
	\bibfield  {author} {\bibinfo {author} {\bibfnamefont {C.~A.}\ \bibnamefont {Ullrich}},\ }\href {https://doi.org/10.1093/acprof:oso/9780199563029.001.0001} {\emph {\bibinfo {title} {Time-Dependent Density-Functional Theory: Concepts and Applications}}}\ (\bibinfo  {publisher} {Oxford University Press, Oxford},\ \bibinfo {year} {2011})\BibitemShut {NoStop}%
	\bibitem [{\citenamefont {Eich}\ and\ \citenamefont {Gross}(2013)}]{Eich2013a}%
	\BibitemOpen
	\bibfield  {author} {\bibinfo {author} {\bibfnamefont {F.~G.}\ \bibnamefont {Eich}}\ and\ \bibinfo {author} {\bibfnamefont {E.~K.~U.}\ \bibnamefont {Gross}},\ }\bibfield  {title} {\bibinfo {title} {Transverse spin-gradient functional for noncollinear spin-density-functional theory},\ }\href {https://doi.org/10.1103/PhysRevLett.111.156401} {\bibfield  {journal} {\bibinfo  {journal} {Phys. Rev. Lett.}\ }\textbf {\bibinfo {volume} {111}},\ \bibinfo {pages} {156401} (\bibinfo {year} {2013})}\BibitemShut {NoStop}%
	\bibitem [{\citenamefont {Egidi}\ \emph {et~al.}(2017)\citenamefont {Egidi}, \citenamefont {Sun}, \citenamefont {Goings}, \citenamefont {Scalmani}, \citenamefont {Frisch},\ and\ \citenamefont {Li}}]{Egidi2017}%
	\BibitemOpen
	\bibfield  {author} {\bibinfo {author} {\bibfnamefont {F.}~\bibnamefont {Egidi}}, \bibinfo {author} {\bibfnamefont {S.}~\bibnamefont {Sun}}, \bibinfo {author} {\bibfnamefont {J.~J.}\ \bibnamefont {Goings}}, \bibinfo {author} {\bibfnamefont {G.}~\bibnamefont {Scalmani}}, \bibinfo {author} {\bibfnamefont {M.~J.}\ \bibnamefont {Frisch}},\ and\ \bibinfo {author} {\bibfnamefont {X.}~\bibnamefont {Li}},\ }\bibfield  {title} {\bibinfo {title} {Two-component noncollinear time-dependent spin density functional theory for excited state calculations},\ }\href {https://doi.org/10.1021/acs.jctc.7b00104} {\bibfield  {journal} {\bibinfo  {journal} {J. Chem. Theory Comput.}\ }\textbf {\bibinfo {volume} {13}},\ \bibinfo {pages} {2591} (\bibinfo {year} {2017})}\BibitemShut {NoStop}%
	\bibitem [{\citenamefont {Lacombe}\ and\ \citenamefont {Maitra}(2023)}]{Lacombe2023}%
	\BibitemOpen
	\bibfield  {author} {\bibinfo {author} {\bibfnamefont {L.}~\bibnamefont {Lacombe}}\ and\ \bibinfo {author} {\bibfnamefont {N.~T.}\ \bibnamefont {Maitra}},\ }\bibfield  {title} {\bibinfo {title} {Non-adiabatic approximations in time-dependent density functional theory: progress and prospects},\ }\href {https://doi.org/10.1038/s41524-023-01061-0} {\bibfield  {journal} {\bibinfo  {journal} {npj Comput. Mater.}\ }\textbf {\bibinfo {volume} {9}},\ \bibinfo {pages} {124} (\bibinfo {year} {2023})}\BibitemShut {NoStop}%
	\bibitem [{\citenamefont {Belashchenko}\ \emph {et~al.}(2016)\citenamefont {Belashchenko}, \citenamefont {Kovalev},\ and\ \citenamefont {van Schilfgaarde}}]{Belashchenko2016}%
	\BibitemOpen
	\bibfield  {author} {\bibinfo {author} {\bibfnamefont {K.~D.}\ \bibnamefont {Belashchenko}}, \bibinfo {author} {\bibfnamefont {A.~A.}\ \bibnamefont {Kovalev}},\ and\ \bibinfo {author} {\bibfnamefont {M.}~\bibnamefont {van Schilfgaarde}},\ }\bibfield  {title} {\bibinfo {title} {Theory of spin loss at metallic interfaces},\ }\href {https://doi.org/10.1103/PhysRevLett.117.207204} {\bibfield  {journal} {\bibinfo  {journal} {Phys. Rev. Lett.}\ }\textbf {\bibinfo {volume} {117}},\ \bibinfo {pages} {207204} (\bibinfo {year} {2016})}\BibitemShut {NoStop}%
	\bibitem [{\citenamefont {Dolui}\ and\ \citenamefont {Nikoli\'{c}}(2017)}]{Dolui2017}%
	\BibitemOpen
	\bibfield  {author} {\bibinfo {author} {\bibfnamefont {K.}~\bibnamefont {Dolui}}\ and\ \bibinfo {author} {\bibfnamefont {B.~K.}\ \bibnamefont {Nikoli\'{c}}},\ }\bibfield  {title} {\bibinfo {title} {Spin-memory loss due to spin-orbit coupling at ferromagnet/heavy-metal interfaces: {\em Ab initio} spin-density matrix approach},\ }\href {https://doi.org/10.1103/PhysRevB.96.220403} {\bibfield  {journal} {\bibinfo  {journal} {Phys. Rev. B}\ }\textbf {\bibinfo {volume} {96}},\ \bibinfo {pages} {220403(R)} (\bibinfo {year} {2017})}\BibitemShut {NoStop}%
	\bibitem [{\citenamefont {Gupta}\ \emph {et~al.}(2020)\citenamefont {Gupta}, \citenamefont {Wesselink}, \citenamefont {Liu}, \citenamefont {Yuan},\ and\ \citenamefont {Kelly}}]{Gupta2020}%
	\BibitemOpen
	\bibfield  {author} {\bibinfo {author} {\bibfnamefont {K.}~\bibnamefont {Gupta}}, \bibinfo {author} {\bibfnamefont {R.~J.~H.}\ \bibnamefont {Wesselink}}, \bibinfo {author} {\bibfnamefont {R.}~\bibnamefont {Liu}}, \bibinfo {author} {\bibfnamefont {Z.}~\bibnamefont {Yuan}},\ and\ \bibinfo {author} {\bibfnamefont {P.~J.}\ \bibnamefont {Kelly}},\ }\bibfield  {title} {\bibinfo {title} {Disorder dependence of interface spin memory loss},\ }\href {https://doi.org/10.1103/PhysRevLett.124.087702} {\bibfield  {journal} {\bibinfo  {journal} {Phys. Rev. Lett.}\ }\textbf {\bibinfo {volume} {124}},\ \bibinfo {pages} {087702} (\bibinfo {year} {2020})}\BibitemShut {NoStop}%
	\bibitem [{\citenamefont {Amin}\ and\ \citenamefont {Stiles}(2016{\natexlab{a}})}]{Amin2016}%
	\BibitemOpen
	\bibfield  {author} {\bibinfo {author} {\bibfnamefont {V.~P.}\ \bibnamefont {Amin}}\ and\ \bibinfo {author} {\bibfnamefont {M.~D.}\ \bibnamefont {Stiles}},\ }\bibfield  {title} {\bibinfo {title} {Spin transport at interfaces with spin-orbit coupling: Formalism},\ }\href {https://doi.org/10.1103/PhysRevB.94.104419} {\bibfield  {journal} {\bibinfo  {journal} {Phys. Rev. B}\ }\textbf {\bibinfo {volume} {94}},\ \bibinfo {pages} {104419} (\bibinfo {year} {2016}{\natexlab{a}})}\BibitemShut {NoStop}%
	\bibitem [{\citenamefont {Amin}\ and\ \citenamefont {Stiles}(2016{\natexlab{b}})}]{Amin2016a}%
	\BibitemOpen
	\bibfield  {author} {\bibinfo {author} {\bibfnamefont {V.~P.}\ \bibnamefont {Amin}}\ and\ \bibinfo {author} {\bibfnamefont {M.~D.}\ \bibnamefont {Stiles}},\ }\bibfield  {title} {\bibinfo {title} {Spin transport at interfaces with spin-orbit coupling: Phenomenology},\ }\href {https://doi.org/10.1103/PhysRevB.94.104420} {\bibfield  {journal} {\bibinfo  {journal} {Phys. Rev. B}\ }\textbf {\bibinfo {volume} {94}},\ \bibinfo {pages} {104420} (\bibinfo {year} {2016}{\natexlab{b}})}\BibitemShut {NoStop}%
	\bibitem [{\citenamefont {Wahada}\ \emph {et~al.}(2022)\citenamefont {Wahada}, \citenamefont {Şaşıoğlu}, \citenamefont {Hoppe}, \citenamefont {Zhou}, \citenamefont {Deniz}, \citenamefont {Rouzegar}, \citenamefont {Kampfrath}, \citenamefont {Mertig}, \citenamefont {Parkin},\ and\ \citenamefont {Woltersdorf}}]{Wahada2022}%
	\BibitemOpen
	\bibfield  {author} {\bibinfo {author} {\bibfnamefont {M.~A.}\ \bibnamefont {Wahada}}, \bibinfo {author} {\bibfnamefont {E.}~\bibnamefont {Şaşıoğlu}}, \bibinfo {author} {\bibfnamefont {W.}~\bibnamefont {Hoppe}}, \bibinfo {author} {\bibfnamefont {X.}~\bibnamefont {Zhou}}, \bibinfo {author} {\bibfnamefont {H.}~\bibnamefont {Deniz}}, \bibinfo {author} {\bibfnamefont {R.}~\bibnamefont {Rouzegar}}, \bibinfo {author} {\bibfnamefont {T.}~\bibnamefont {Kampfrath}}, \bibinfo {author} {\bibfnamefont {I.}~\bibnamefont {Mertig}}, \bibinfo {author} {\bibfnamefont {S.~S.~P.}\ \bibnamefont {Parkin}},\ and\ \bibinfo {author} {\bibfnamefont {G.}~\bibnamefont {Woltersdorf}},\ }\bibfield  {title} {\bibinfo {title} {Atomic scale control of spin current transmission at interfaces},\ }\href {https://doi.org/10.1021/acs.nanolett.1c04358} {\bibfield  {journal} {\bibinfo  {journal} {Nano Lett.}\ }\textbf {\bibinfo {volume} {22}},\ \bibinfo {pages} {3539} (\bibinfo {year} {2022})}\BibitemShut {NoStop}%
	\bibitem [{\citenamefont {Van~Tuan}\ \emph {et~al.}(2016)\citenamefont {Van~Tuan}, \citenamefont {Marmolejo-Tejada}, \citenamefont {Waintal}, \citenamefont {Nikoli\'{c}}, \citenamefont {Valenzuela},\ and\ \citenamefont {Roche}}]{VanTuan2016}%
	\BibitemOpen
	\bibfield  {author} {\bibinfo {author} {\bibfnamefont {D.}~\bibnamefont {Van~Tuan}}, \bibinfo {author} {\bibfnamefont {J.~M.}\ \bibnamefont {Marmolejo-Tejada}}, \bibinfo {author} {\bibfnamefont {X.}~\bibnamefont {Waintal}}, \bibinfo {author} {\bibfnamefont {B.~K.}\ \bibnamefont {Nikoli\'{c}}}, \bibinfo {author} {\bibfnamefont {S.~O.}\ \bibnamefont {Valenzuela}},\ and\ \bibinfo {author} {\bibfnamefont {S.}~\bibnamefont {Roche}},\ }\bibfield  {title} {\bibinfo {title} {Spin hall effect and origins of nonlocal resistance in adatom-decorated graphene},\ }\href {https://doi.org/10.1103/PhysRevLett.117.176602} {\bibfield  {journal} {\bibinfo  {journal} {Phys. Rev. Lett.}\ }\textbf {\bibinfo {volume} {117}},\ \bibinfo {pages} {176602} (\bibinfo {year} {2016})}\BibitemShut {NoStop}%
	\bibitem [{\citenamefont {Chu}\ \emph {et~al.}(2024)\citenamefont {Chu}, \citenamefont {Yang}, \citenamefont {Li}, \citenamefont {Hwangbo}, \citenamefont {Wen}, \citenamefont {Bielinski}, \citenamefont {Zhang}, \citenamefont {Martinson}, \citenamefont {Hruszkewycz}, \citenamefont {Fong} \emph {et~al.}}]{Chu2024}%
	\BibitemOpen
	\bibfield  {author} {\bibinfo {author} {\bibfnamefont {Z.}~\bibnamefont {Chu}}, \bibinfo {author} {\bibfnamefont {J.}~\bibnamefont {Yang}}, \bibinfo {author} {\bibfnamefont {Y.}~\bibnamefont {Li}}, \bibinfo {author} {\bibfnamefont {K.}~\bibnamefont {Hwangbo}}, \bibinfo {author} {\bibfnamefont {J.}~\bibnamefont {Wen}}, \bibinfo {author} {\bibfnamefont {A.~R.}\ \bibnamefont {Bielinski}}, \bibinfo {author} {\bibfnamefont {Q.}~\bibnamefont {Zhang}}, \bibinfo {author} {\bibfnamefont {A.~B.~F.}\ \bibnamefont {Martinson}}, \bibinfo {author} {\bibfnamefont {S.~O.}\ \bibnamefont {Hruszkewycz}}, \bibinfo {author} {\bibfnamefont {D.~D.}\ \bibnamefont {Fong}}, \emph {et~al.},\ }\bibfield  {title} {\bibinfo {title} {Revealing subterahertz atomic vibrations in quantum paraelectrics by surface-sensitive spintronic terahertz spectroscopy},\ }\href {https://doi.org/10.1126/sciadv.ads8601} {\bibfield  {journal} {\bibinfo  {journal} {Sci. Adv.}\ }\textbf {\bibinfo {volume} {10}},\ \bibinfo {pages} {eads8601} (\bibinfo {year}
		{2024})}\BibitemShut {NoStop}%
	\bibitem [{\citenamefont {Stefanucci}\ and\ \citenamefont {van Leeuwen}(2025)}]{Stefanucci2025}%
	\BibitemOpen
	\bibfield  {author} {\bibinfo {author} {\bibfnamefont {G.}~\bibnamefont {Stefanucci}}\ and\ \bibinfo {author} {\bibfnamefont {R.}~\bibnamefont {van Leeuwen}},\ }\href {https://doi.org/10.1017/CBO9781139023979} {\emph {\bibinfo {title} {Nonequilibrium Many-Body Theory of Quantum Systems: A Modern Introduction}}}\ (\bibinfo  {publisher} {Cambridge University Press, Cambridge},\ \bibinfo {year} {2025})\BibitemShut {NoStop}%
	\bibitem [{\citenamefont {Switkes}\ \emph {et~al.}(1999)\citenamefont {Switkes}, \citenamefont {Marcus}, \citenamefont {Campman},\ and\ \citenamefont {Gossard}}]{Switkes1999}%
	\BibitemOpen
	\bibfield  {author} {\bibinfo {author} {\bibfnamefont {M.}~\bibnamefont {Switkes}}, \bibinfo {author} {\bibfnamefont {C.~M.}\ \bibnamefont {Marcus}}, \bibinfo {author} {\bibfnamefont {K.}~\bibnamefont {Campman}},\ and\ \bibinfo {author} {\bibfnamefont {A.~C.}\ \bibnamefont {Gossard}},\ }\bibfield  {title} {\bibinfo {title} {An adiabatic quantum electron pump},\ }\href {https://doi.org/10.1126/science.283.5409.1905} {\bibfield  {journal} {\bibinfo  {journal} {Science}\ }\textbf {\bibinfo {volume} {283}},\ \bibinfo {pages} {1905} (\bibinfo {year} {1999})}\BibitemShut {NoStop}%
	\bibitem [{\citenamefont {Petrovi\ifmmode~\acute{c}\else \'{c}\fi{}}\ \emph {et~al.}(2018)\citenamefont {Petrovi\ifmmode~\acute{c}\else \'{c}\fi{}}, \citenamefont {Popescu}, \citenamefont {Bajpai}, \citenamefont {Plech\'a\ifmmode~\check{c}\else \v{c}\fi{}},\ and\ \citenamefont {Nikoli\ifmmode~\acute{c}\else \'{c}\fi{}}}]{Petrovic2018}%
	\BibitemOpen
	\bibfield  {author} {\bibinfo {author} {\bibfnamefont {M.~D.}\ \bibnamefont {Petrovi\ifmmode~\acute{c}\else \'{c}\fi{}}}, \bibinfo {author} {\bibfnamefont {B.~S.}\ \bibnamefont {Popescu}}, \bibinfo {author} {\bibfnamefont {U.}~\bibnamefont {Bajpai}}, \bibinfo {author} {\bibfnamefont {P.}~\bibnamefont {Plech\'a\ifmmode~\check{c}\else \v{c}\fi{}}},\ and\ \bibinfo {author} {\bibfnamefont {B.~K.}\ \bibnamefont {Nikoli\ifmmode~\acute{c}\else \'{c}\fi{}}},\ }\bibfield  {title} {\bibinfo {title} {Spin and charge pumping by a steady or pulse-current-driven magnetic domain wall: A self-consistent multiscale time-dependent quantum-classical hybrid approach},\ }\href {https://doi.org/10.1103/PhysRevApplied.10.054038} {\bibfield  {journal} {\bibinfo  {journal} {Phys. Rev. Applied}\ }\textbf {\bibinfo {volume} {10}},\ \bibinfo {pages} {054038} (\bibinfo {year} {2018})}\BibitemShut {NoStop}%
	\bibitem [{\citenamefont {Petrovi\ifmmode~\acute{c}\else \'{c}\fi{}}\ \emph {et~al.}(2021)\citenamefont {Petrovi\ifmmode~\acute{c}\else \'{c}\fi{}}, \citenamefont {Bajpai}, \citenamefont {Plech\'a\ifmmode~\check{c}\else \v{c}\fi{}},\ and\ \citenamefont {Nikoli\ifmmode~\acute{c}\else \'{c}\fi{}}}]{Petrovic2021}%
	\BibitemOpen
	\bibfield  {author} {\bibinfo {author} {\bibfnamefont {M.~D.}\ \bibnamefont {Petrovi\ifmmode~\acute{c}\else \'{c}\fi{}}}, \bibinfo {author} {\bibfnamefont {U.}~\bibnamefont {Bajpai}}, \bibinfo {author} {\bibfnamefont {P.}~\bibnamefont {Plech\'a\ifmmode~\check{c}\else \v{c}\fi{}}},\ and\ \bibinfo {author} {\bibfnamefont {B.~K.}\ \bibnamefont {Nikoli\ifmmode~\acute{c}\else \'{c}\fi{}}},\ }\bibfield  {title} {\bibinfo {title} {Annihilation of topological solitons in magnetism with spin-wave burst finale: {Role} of nonequilibrium electrons causing nonlocal damping and spin pumping over ultrabroadband frequency range},\ }\href {https://doi.org/10.1103/PhysRevB.104.L020407} {\bibfield  {journal} {\bibinfo  {journal} {Phys. Rev. B}\ }\textbf {\bibinfo {volume} {104}},\ \bibinfo {pages} {L020407} (\bibinfo {year} {2021})}\BibitemShut {NoStop}%
	\bibitem [{\citenamefont {Abbout}\ \emph {et~al.}(2018)\citenamefont {Abbout}, \citenamefont {Weston}, \citenamefont {Waintal},\ and\ \citenamefont {Manchon}}]{Abbout2018}%
	\BibitemOpen
	\bibfield  {author} {\bibinfo {author} {\bibfnamefont {A.}~\bibnamefont {Abbout}}, \bibinfo {author} {\bibfnamefont {J.}~\bibnamefont {Weston}}, \bibinfo {author} {\bibfnamefont {X.}~\bibnamefont {Waintal}},\ and\ \bibinfo {author} {\bibfnamefont {A.}~\bibnamefont {Manchon}},\ }\bibfield  {title} {\bibinfo {title} {Cooperative charge pumping and enhanced skyrmion mobility},\ }\href {https://doi.org/10.1103/PhysRevLett.121.257203} {\bibfield  {journal} {\bibinfo  {journal} {Phys. Rev. Lett.}\ }\textbf {\bibinfo {volume} {121}},\ \bibinfo {pages} {257203} (\bibinfo {year} {2018})}\BibitemShut {NoStop}%
	\bibitem [{\citenamefont {Chen}\ \emph {et~al.}(2009)\citenamefont {Chen}, \citenamefont {Chang}, \citenamefont {Xiao},\ and\ \citenamefont {Nikoli\ifmmode~\acute{c}\else \'{c}\fi{}}}]{Chen2009}%
	\BibitemOpen
	\bibfield  {author} {\bibinfo {author} {\bibfnamefont {S.-H.}\ \bibnamefont {Chen}}, \bibinfo {author} {\bibfnamefont {C.-R.}\ \bibnamefont {Chang}}, \bibinfo {author} {\bibfnamefont {J.~Q.}\ \bibnamefont {Xiao}},\ and\ \bibinfo {author} {\bibfnamefont {B.~K.}\ \bibnamefont {Nikoli\ifmmode~\acute{c}\else \'{c}\fi{}}},\ }\bibfield  {title} {\bibinfo {title} {Spin and charge pumping in magnetic tunnel junctions with precessing magnetization: A nonequilibrium green function approach},\ }\href {https://doi.org/10.1103/PhysRevB.79.054424} {\bibfield  {journal} {\bibinfo  {journal} {Phys. Rev. B}\ }\textbf {\bibinfo {volume} {79}},\ \bibinfo {pages} {054424} (\bibinfo {year} {2009})}\BibitemShut {NoStop}%
\end{thebibliography}
\end{document}